\documentclass{ieeeaccess}
\usepackage{cite}
\usepackage{amsmath,amssymb,amsfonts}

\usepackage{graphicx}
\usepackage{textcomp}

\usepackage{bm}
\makeatletter
\AtBeginDocument{\DeclareMathVersion{bold}
\SetSymbolFont{operators}{bold}{T1}{times}{b}{n}
\SetSymbolFont{NewLetters}{bold}{T1}{times}{b}{it}
\SetMathAlphabet{\mathrm}{bold}{T1}{times}{b}{n}
\SetMathAlphabet{\mathit}{bold}{T1}{times}{b}{it}
\SetMathAlphabet{\mathbf}{bold}{T1}{times}{b}{n}
\SetMathAlphabet{\mathtt}{bold}{OT1}{pcr}{b}{n}
\SetSymbolFont{symbols}{bold}{OMS}{cmsy}{b}{n}
\renewcommand\boldmath{\@nomath\boldmath\mathversion{bold}}}
\makeatother

\def\BibTeX{{\rm B\kern-.05em{\sc i\kern-.025em b}\kern-.08em
    T\kern-.1667em\lower.7ex\hbox{E}\kern-.125emX}}

\ExplSyntaxOn\cs_set_eq:NN\ManulyxClassYear\year\cs_set_eq:NN\year\tex_year:D\ExplSyntaxOff
\usepackage{etoolbox}
\usepackage{xspace}
\usepackage{booktabs,tabularx,array}
\usepackage{siunitx}
\usepackage[font=small,labelfont=bf]{caption}

\usepackage{enumitem}
\usepackage{xcolor}
\usepackage{placeins}
\usepackage{tikz}
\usetikzlibrary{arrows.meta,calc}
\usepackage{algorithm}
\usepackage{algorithmicx}
\usepackage{algpseudocode}
\usepackage{float}
\usepackage{array}
\usepackage[caption=false,font=normalsize,labelfont=sf,textfont=sf]{subfig}
\usepackage{stfloats}
\usepackage{url}
\usepackage{verbatim}
\usepackage{indentfirst}
\usepackage{cases}
\definecolor{paperblue}{RGB}{30,90,160}
\usepackage{acronym}

\acrodef{abl}[ABL]{atmospheric boundary layer}
\acrodef{afgl}[AFGL]{Air Force Geophysics Laboratory}
\acrodef{agl}[AGL]{above ground level}
\acrodef{bl}[BL]{boundary-layer}
\acrodef{cbl}[CBL]{convective boundary layer}
\acrodef{dimm}[DIMM]{Differential Image Motion Monitor}
\acrodef{hv}[HV]{Hufnagel--Valley}
\acrodef{mass}[MASS]{Multi-Aperture Scintillation Sensor}
\acrodef{most}[MOST]{Monin--Obukhov similarity theory}
\acrodef{msl}[MSL]{mean sea level}
\acrodef{ri}[RI]{refractive-index}
\acrodef{rms}[rms]{root-mean-square}
\acrodef{scidar}[SCIDAR]{scintillation detection and ranging}
\acrodef{sl}[SL]{surface-layer}

\providecommand{\Cn}{}\renewcommand{\Cn}{C_n^2}
\providecommand{\CT}{}\renewcommand{\CT}{C_T^2}
\providecommand{\Cth}{}\renewcommand{\Cth}{C_\theta^2}
\providecommand{\Cq}{}\renewcommand{\Cq}{C_q^2}

\def\tsc#1{\csdef{#1}{\textsc{\lowercase{#1}}\xspace}}
\tsc{WGM}
\tsc{QE}
\tsc{EP}
\tsc{PMS}
\tsc{BEC}
\tsc{DE}
\definecolor{accessblue}{cmyk}{1,0.3,0,0.2}

\let\year\ManulyxClassYear
\providecommand{\citep}{\cite}

\makeatletter
\def\headerlogo{}\def\headerlogoall{}\def\headName{}
\def\doifont#1\vss{\vss}
\def\history#1{}\let\@history\@empty
\def\ps@titlepage{\let\@oddhead\@empty\let\@evenhead\@empty\def\@oddfoot{\hfil\thepage\hfil}\let\@evenfoot\@oddfoot}
\def\ps@headings{\def\@oddhead{\hfill{\headerfont\rightmark}}\def\@evenhead{{\headerfont\leftmark}\hfill}
  \def\@oddfoot{\hfil\thepage\hfil}\let\@evenfoot\@oddfoot}
\makeatother
\ExplSyntaxOn
\cs_if_exist:cT { includegraphics } {
  \exp_args:NNc \NewCommandCopy \mxpreprintpicture { includegraphics }
  \exp_args:Nc \RenewDocumentCommand { includegraphics } { s O{} m } {
    \str_set:Nx \l_tmpa_str { \str_foldcase:n { #3 } }
    \str_if_in:NnTF \l_tmpa_str { orcid } { \mx_preprint_draw:nnn {#1} {#2} {#3} } {
      \str_if_in:NnTF \l_tmpa_str { logo } { \phantom { \mx_preprint_draw:nnn {#1} {#2} {#3} } } {
        \str_if_in:NnTF \l_tmpa_str { crossmark } { \phantom { \mx_preprint_draw:nnn {#1} {#2} {#3} } } {
          \mx_preprint_draw:nnn {#1} {#2} {#3} } } } } }
\cs_new:Npn \mx_preprint_draw:nnn #1#2#3 { \IfBooleanTF {#1} { \mxpreprintpicture* [#2] {#3} } { \mxpreprintpicture [#2] {#3} } }
\ExplSyntaxOff
\begin{document}

\history{Date of publication xxxx 00, 0000, date of current version xxxx 00, 0000.}
\doi{10.1109/ACCESS.2024.0429000}

\title{A regime dependent parameterization of the refractive-index structure parameter $C_n^2$}
\author{Yalçın~Ata\authorrefmark{1,2},
        Muhsin~Caner~Gökçe\authorrefmark{3,4},
        and Yahya~Baykal\authorrefmark{5}}
\address[1]{Department of Electrical and Electronics Engineering, OSTIM Technical University, OSTIM Yenimahalle, Ankara, 06374, Türkiye (e-mail: ylcnata@gmail.com)}
\address[2]{Scopt Inc., One Boston Place Suite 2600, Boston, 02108, Massachussets, United States}
\address[3]{TED University, Ziya Gökalp Caddesi No: 47 - 48, 06420, Ankara, Türkiye}
\address[4]{Delft University of Technology, 2629 HS, Delft, The Netherlands (e-mail: M.C.Gokce@tudelft.nl)}
\address[5]{Çankaya University, Yukarıyurtçu mah. Mimar Sinan cad. No. 4, Etimesgut, 06815, Ankara, Türkiye (e-mail: y.baykal@cankaya.edu.tr)}

\markboth{IEEE Access,~Vol.~XX, No.~X, Month~2026}
{Ata \MakeLowercase{\textit{et al.}}: A regime dependent parameterization of the refractive-index structure parameter $C_n^2$}

\corresp{Corresponding author: Muhsin Caner Gökçe (e-mail: muhsin.gokce@tedu.edu.tr).}

\begin{abstract}
Vertical profiles of the refractive-index (RI) structure parameter, $C_n^2$, are required in atmospheric-optics calculations. The Hufnagel--Valley (HV) model is a useful reference profile and it is widely used however, it is difficult to say that it produces consistent results at the observed altitude, magnitude, and thickness of all turbulent layers and it can diverge from realistic measurements in particular height ranges, sites, and stability regimes. This study develops a regime dependent RI structure parameter considering meteorological conditions and, more importantly, it provides a structure that can be calibrated against measured $C_n^2$ throughout the surface layer (SL), atmospheric boundary layer (ABL), and free atmosphere. The formulation combines SL including distinct unstable, near-neutral, and stable limits; convective ABL; and free-atmospheric turbulence when kinetic-energy and temperature-variance dissipation rates are available. The $C_n^2$ is formed by weighting the contribution of different layers. We use representative, physically plausible meteorological profiles to examine the model’s limiting behavior, transitions between atmospheric regimes, continuity, sensitivity to its parameters, and propagation-related differences from the HV 5/7 model The intended outcome is a measurement-calibrated model that remains physically interpretable across meteorological regimes.
\end{abstract}


\titlepgskip=-21pt

\maketitle

\acresetall

\section{Introduction}

Random \ac{ri} fluctuations perturb the phase and amplitude of an optical wave as it propagates through the atmosphere. Their consequences include loss of spatial coherence, image blur, beam spreading and wander, angle-of-arrival fluctuations, and irradiance scintillation. These effects constrain astronomical imaging, free-space optical links, surveillance, tracking, and directed-energy systems \citep{tatarskii1961wave,hufnagel1978propagation,beland1993propagation}. In the inertial subrange, the conventional measure of turbulence strength is the \ac{ri} structure parameter $\Cn$. Its vertical distribution matters as much as its column integral: the Fried coherence diameter mainly weights $\Cn$ uniformly along a vertical path, the isoplanatic angle weights high-altitude turbulence by approximately $z^{5/3}$, and the Greenwood frequency introduces an additional $V^{5/3}$ transverse-wind weighting \citep{fried1965statistics,greenwood1977bandwidth,beland1993propagation}. A profile that reproduces a near-ground observation can therefore still give an incorrect adaptive-optics or propagation prediction if it places turbulence at the wrong altitude.

The difficulty is that $\Cn$ is not an altitude-only state variable. It is a statistical property of \ac{ri} fluctuations and consequently depends on the production, transport, and dissipation of thermodynamic variance. In the \ac{sl}, heat and moisture fluxes, friction velocity, stability, roughness, and displacement height control the scalar structure parameters \citep{wyngaard1971behavior,andreas1988estimating,tunick2003cn2}. In a daytime \ac{cbl}, vigorous mixing produces a nearly uniform mean potential-temperature profile while turbulent temperature variance remains nonzero; a model based only on the local mean gradient can therefore predict an artificial minimum \citep{kaimal1976turbulence,wyngaard1980behavior,fairall1987top}. Above the \ac{bl}, optical turbulence is often organized into intermittent sheets associated with shear, inversions, gravity waves, and the tropopause rather than into a smooth monotonic background \citep{coulman1988outer,muschinski2024anisotropy,muschinski2025predicting}. These mechanisms motivate a regime-dependent treatment for $C_n^2$ representation.

The limitation of \ac{hv} is therefore broader than the absence of explicit meteorological inputs. Its smooth terms prescribe the altitude of the ground-layer decay, the background decrease, and the upper-atmospheric maximum in advance. Measurements, however, show site-dependent inversion and turbulent-layer heights, strong diurnal changes, and localized layers that cannot in general be moved or reshaped by adjusting only $A$ (nominal value of $C_n^2$ at ground level in \si{m^{-2/3}}) and $w$ (wind speed) \citep{good1988atmospheric,tofsted2006atmospheric,quatresooz2023computation}. Consequently, a model may correctly predict integrated seeing while failing to reproduce the measured $C_n^2$ profile within certain altitude ranges. This discrepancy is important because different optical observables assign different weights to turbulence at different altitudes.

The aim of this paper is to construct and examine a full-column $\Cn$ parameterization in which the governing expression changes with the diagnosed turbulence regime and the available meteorological information. The long-term objective is a model that is quantitatively consistent with measured profiles in convective, neutral, stable, transition, and free-atmospheric conditions, rather than merely a model that contains more meteorological variables. We do not claim that the present synthetic calculations have already achieved that objective. Instead, we formulate a physically traceable hypothesis, identify which equations are inherited from the literature, state which connections are introduced here, and define how those connections should be calibrated and independently validated against measurements.

\subsection{Related works}

The historical development of the \ac{hv} model helps explain its three-term structure. Hufnagel's original upper-air expression contained two terms: a wind-dependent high-altitude term and a background exponential term; it was intended for use above the lower atmosphere \citep{hufnagel1978propagation}. Beland reports that Ulrich subsequently extended this expression to the ground, following Valley's suggestion, by adding an exponential \ac{bl} term \citep{valley1980isoplanatic,ulrich1988hufnagel,beland1993propagation}. This extension became known as the \ac{hv} model.

Existing $\Cn$ models reflect a trade-off between input requirements and meteorological specificity. The \ac{hv} family occupies one end of this trade-off. It is analytic, inexpensive, and useful when only a representative vertical profile is needed \citep{hufnagel1978propagation,good1988atmospheric,beland1993propagation}. Its three terms represent a ground contribution, a background tropospheric contribution, and a wind-dependent upper-level contribution. Good et al. emphasized, however, that atmospheric models are averaged representations and may not describe a particular realization \citep{good1988atmospheric}. The standard \ac{hv} 5/7 settings are especially associated with typical nighttime astronomical values, whereas Tofsted noted that the same profile is less suitable for daytime \ac{bl} conditions and elevated inversion structure \citep{tofsted2006atmospheric}.

Meteorology-based surface models take a different approach. The Kansas experiments established similarity forms for temperature and \ac{ri} structure parameters under unstable, neutral, and stable stratification \citep{wyngaard1971behavior,businger1971flux}. Andreas developed a wavelength-aware conversion from temperature and humidity statistics to $\Cn$ and demonstrated how routine meteorological quantities can be used over snow and sea ice \citep{andreas1988estimating}. Tunick subsequently implemented a semi-empirical \ac{sl} model driven by two-level wind, temperature, and humidity observations and compared its predictions with scintillometer measurements \citep{tunick2003cn2}. These studies show why a fixed coefficient $A$ cannot represent the full range of surface states: the same height can experience orders-of-magnitude variation as the surface flux and stability change.

Models for the remaining atmospheric \ac{bl} must account for vertical organization that is absent from a surface law. Kaimal et al. documented the three-part behavior of scalar structure in a \ac{cbl}: a free-convective decrease, a weakly varying mixed-layer interior, and enhancement near the capping inversion \citep{kaimal1976turbulence}. Wyngaard and LeMone and Fairall related the upper-layer behavior to entrainment and top-down as well as bottom-up scalar transport \citep{wyngaard1980behavior,fairall1987top}. Composite engineering profiles reviewed by Good et al. and developed further by Tofsted combine such \ac{bl} behavior with surface and upper-tropospheric models \citep{good1988atmospheric,tofsted2006atmospheric}. Their work directly motivates the regional architecture adopted here, while also showing that the location and amplitude of the entrainment peak are site- and state-dependent.

In the free atmosphere, a common family of methods combines a \ac{ri} gradient with a prescribed or diagnosed outer scale \citep{coulman1988outer,good1988atmospheric,beland1993propagation}. Such methods can be evaluated from radiosondes, and recent work has used Thorpe, Ellison, and statistical increment scales to recover site-specific turbulent layers \citep{wu2020simple,quatresooz2023computation}. Their main limitation is not simply measurement noise: outer-scale definitions are not interchangeable, and gradient estimates are sensitive to vertical sampling and preprocessing \citep{muschinski2024anisotropy}. A second family obtains thermodynamic structure parameters from turbulence closures in numerical weather prediction. Wang et al. predicted temperature, humidity, and cross-structure parameters with a second-order closure throughout a convective marine \ac{bl} \citep{wang2020computing}. Muschinski et al. used kinetic-energy and temperature-variance dissipation rates in a general circulation model to estimate the large-scale free-atmospheric distribution of $\Cn$, while explicitly recognizing that a roughly \SI{500}{m} vertical grid cannot resolve thin turbulent sheets \citep{muschinski2025predicting}.

The literature therefore supplies several credible component models, but it does not provide a unique rule for joining them. Surface similarity should not be extended through the whole \ac{bl}; a local gradient closure is unreliable in the well-mixed convective interior; and a dissipation closure requires variables that are seldom available from routine stations. Our contribution addresses this practical gap. We retain the physical domain of each component, introduce an explicit and auditable transition rule, and define reduced modes for cases in which only surface observations or radiosonde profiles are available. The resulting construction is closer in spirit to a model hierarchy than to a new universal altitude curve.

The literature provides several physically meaningful models for different parts of the atmosphere, but there is no generally accepted method for combining them into a single continuous $C_n^2$ profile. \Ac{sl} similarity relations should not be applied throughout the entire \ac{abl}, local gradient-based models may perform poorly in the well-mixed interior of a \ac{cbl}, and dissipation-based models require variables that are not usually available from routine meteorological stations. The present study addresses this problem by applying each component model only within the atmospheric region for which it is physically appropriate and by connecting the components through clearly defined transition functions. We also introduce simplified operating modes for situations in which only surface observations or radiosonde profiles are available. Thus, the proposed formulation should be viewed as a flexible hierarchy of models rather than as a single universal function of altitude.

\subsection{Motivation and contribution}

The central hypothesis of this study is that a composite profile conditioned on observable meteorological regimes can preserve the practical advantages of \ac{hv} while correcting two different limitations: the absence of state dependence and the inability of a fixed smooth profile to follow measured altitude-dependent structure. The main contribution of this study is a reproducible and flexible modeling hierarchy that extends beyond a single altitude-dependent turbulence law. The proposed framework connects established component relations, clearly identifies the atmospheric regions in which they are applicable, and distinguishes the physical relations adopted from the literature from the numerical and blending choices introduced in this work. The formulation has four objectives:
\begin{enumerate}[leftmargin=2em,rightmargin=2pt]
\item replace the fixed \ac{hv} ground exponential with meteorology responsive surface physics;
\item retain distinct convective, stable/neutral, and free atmospheric mechanisms; and
\item permit the heights, amplitudes, and thicknesses of turbulent layers to respond to measured or modeled atmospheric structure rather than remain fixed by climatology;
\item provide parameters that can be calibrated by altitude and regime, while degrading transparently to simpler modes when required meteorological data are unavailable.
\end{enumerate}

The specific methodological advances are: a wavelength-dependent refractivity conversion used consistently across all dry-air branches; adoption of a stable similarity relation with the correct $z$-less asymptote; an explicit warning that the two Tatarskii outer-scale conventions cannot be mixed; a dissipation-based preferred branch for the free atmosphere; and a convex blending rule that is positive and continuous by construction. The prescribed-input calculations address three preliminary questions: (i) how strongly the proposed surface term responds to stability and forcing relative to \ac{hv} 5/7, (ii) whether the full-column construction remains continuous while \ac{bl} features move with $h_i$, and (iii) how structural differences affect propagation-weighted turbulence integrals. A fourth and decisive question -- whether the model reduces altitude-resolved error relative to \ac{hv} in every regime -- is reserved for observational calibration and independent validation.
\\
\section{Reference Models and Coordinate Definitions}

\subsection{Hufnagel--Valley reference model and coordinates}

\begin{figure}[t]
\centering
\begin{tikzpicture}[x=0.70cm,y=0.70cm,>=Latex,font=\scriptsize]
  \draw[paperblue,very thick] (0,0) -- (10,0);
  \node[black,anchor=south west] at (7,0.06) {\acs{msl} datum};

  \fill[gray!18] (4.2,1.55) -- (5.4,1.78) -- (6.5,1.62) -- (7.8,1.78)
      -- (10,1.62) -- (10,1.30) -- (4.2,1.30) -- cycle;
  \draw[black,thick] (4.2,1.55) -- (5.4,1.78) -- (6.5,1.62) -- (7.8,1.78) -- (10,1.62);

  \coordinate (site) at (7.0,1.68);
  \coordinate (point) at (7.0,5.10);
  \fill (site) circle (2pt);
  \draw[black,thick] (6.75,1.68) rectangle (7.25,2.08);
  \draw[black] (6.82,2.08) -- (7.0,2.35) -- (7.18,2.08);
  \node[anchor=west] at (7.35,1.93) {site};
  \fill[red!70!black] (point) circle (2.2pt);
  \node[anchor=west] at (7.20,5.10) {point};
  \draw[densely dashed,gray] (point) -- (2.7,5.10);
  \draw[densely dashed,gray] (site) -- (2.7,1.68);

  \draw[<->,very thick] (1.0,0.06) -- (1.0,5.04)
      node[midway,left] {$H$};
  \draw[<->,thick] (2.7,0.06) -- (2.7,1.62)
      node[midway,left] {$H_{\rm site}$};
  \draw[<->,thick] (4.5,1.74) -- (4.5,5.04)
      node[midway,left] {$z$};

  \node[black,fill=white,inner sep=3pt]
      at (8.35,3.75) {$H=H_{\rm site}+z$};
\end{tikzpicture}
\caption{Altitude and local-height coordinates, with $H=H_{\rm site}+z$.}
\label{fig:height_coordinates}
\end{figure}
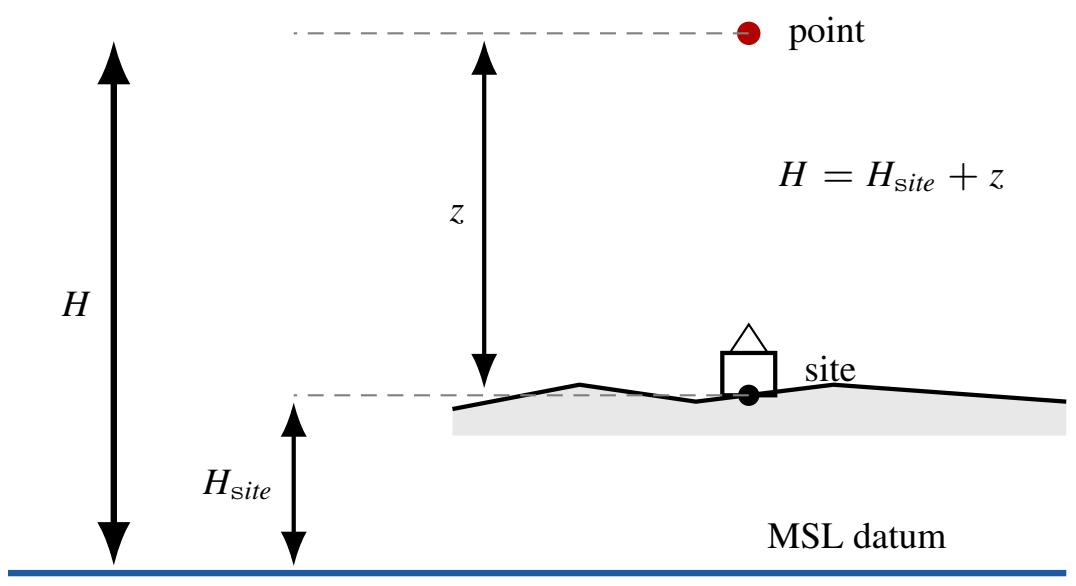

Let $z$ denote geometric \ac{agl}, and let $H=H_{\rm site}+z$ denote altitude \ac{msl}. Fig.~\ref{fig:height_coordinates} illustrates the distinction between these coordinates.

To avoid concealing a coordinate choice, the \ac{hv} comparator in this work follows the explicitly \ac{agl} presentation used by Muschinski et al. \citep[Eq.~(1)]{muschinski2025predicting}:
\begin{equation}
\begin{split}
C_{n,\rm HV}^{2}(z)=& 0.00594\left(\frac{w}{27}\right)^2(10^{-5}z)^{10}e^{-z/1000}\\
&+2.7 \times 10^{-16}e^{-z/1500}+Ae^{-z/100},
\end{split}
\label{eq:hv}
\end{equation}
where $w$ is a \ac{rms} wind parameter in \si{m.s^{-1}}, $A$ is the nominal ground coefficient in \si{m^{-2/3}}, and all heights in the exponentials are in metres. Eq.~\eqref{eq:hv} is the comparator called \ac{hv} 5/7 when $w=\SI{21}{m.s^{-1}}$ and $A=1.7\times10^{-14}\ \si{m^{-2/3}}$ \citep{good1988atmospheric,beland1993propagation}. Hufnagel's earlier upper-atmosphere model and later extensions do not use uniform coordinate descriptions across all sources \citep{hufnagel1978propagation,good1988atmospheric,beland1993propagation}; therefore an \ac{msl} implementation must be declared separately and must not be mixed silently with Eq.~\eqref{eq:hv}.

The wind parameter is commonly written as the squared-wind average printed by Good et al. and Beland \citep[Eq.~(21)]{good1988atmospheric,beland1993propagation}:
\begin{equation}
w=\left[\frac{1}{15000}\int_{5000}^{20000}V^2(z)\,\mathrm{d}z\right]^{1/2},
\label{eq:rmswind}
\end{equation}
where $V$ is transverse wind speed in \si{m.s^{-1}} and the integration limits are metres.

\subsection{Thermodynamic conversion}

The inertial-subrange definition of $\Cn$ is \citep{tatarskii1961wave,beland1993propagation}
\begin{equation}
D_n(r)=\left\langle[n(\bm{x}+\bm{r})-n(\bm{x})]^2\right\rangle=\Cn r^{2/3},
\label{eq:structure}
\end{equation}
where $D_n$ is the \ac{ri} structure function, $n(.)$ is the \ac{ri} and $r$ is separation distance.

For dry air and negligible turbulent pressure fluctuations, Muschinski et al. give \citep[Eqs.~(2)--(6)]{muschinski2024anisotropy}
\begin{align}
n&=1+a\left(1+\frac{b}{\lambda^2}\right)\frac{p}{T},
\label{eq:refractivity}\\
\Cn&=K_n(p,T,\lambda)\CT,
\label{eq:dryconvert}\\
K_n(p,T,\lambda)&=a^2\left(1+\frac{b}{\lambda^2}\right)^2\frac{\langle p \rangle^2}{\langle T\rangle^4},
\label{eq:Kn}
\end{align}
where $a=7.76\times10^{-7}\ \si{K.Pa^{-1}}$, $b=7.52\times10^{-15}\ \si{m^2}$, wavelength $\lambda$ is in metres, pressure $p$ is in pascals, and absolute temperature $T$ is in kelvin. The familiar visible-wavelength approximation $(79\times10^{-6}P/T^2)^2\CT$, with $P$ in hectopascals, is retained only as an approximation near visible wavelengths \citep{andreas1988estimating}.

For moist air, the corresponding first-order structure-parameter relation is \citep{andreas1988estimating,wang2020computing}
\begin{equation}
\Cn=n_T^2\CT+n_q^2\Cq+2n_Tn_qC_{Tq},
\label{eq:moistcn}
\end{equation}
where $n_T=\partial n/\partial T$, $n_q=\partial n/\partial q$, $q$ is specific humidity, and $C_{Tq}$ is the temperature--humidity cross-structure parameter. Eq.~\eqref{eq:moistcn} is the required general form for humid, marine, and infrared applications. 

\section{Regimes and characteristics}

To account for changes in the dominant turbulence-generating processes with altitude, the formulation is organized into three overlapping atmospheric regions. Within the \ac{sl}, $C_n^2$ is estimated from local flux--gradient relationships. In the remainder of the \ac{abl}, the model transitions continuously between formulations appropriate to strongly convective and stable or near-neutral conditions. In the free atmosphere, $C_n^2$ is evaluated using a dissipation-based closure when the required turbulence prognostics are available. This partition is conceptual rather than a set of sharp physical boundaries: adjacent formulations overlap, and smooth membership functions based on stability and height provide continuous transitions between them.

\subsection{Surface layer}

The \ac{sl} is the lowest portion of the \ac{abl} in which turbulent fluxes are approximately height invariant and \ac{most} is applicable. This assumption requires a horizontally homogeneous fetch, approximate stationarity, and evaluation above the roughness sublayer. It becomes questionable close to large obstacles, within forest or urban canopies, during rapid morning and evening transitions, and under very stable intermittent turbulence. These restrictions are physical limitations of the similarity framework rather than numerical details.

Define the aerodynamic height as
\begin{equation}
z_a=z-d,
\label{eq:za}
\end{equation}
where $d$ is displacement height. Eq.~\eqref{eq:za} is a coordinate definition proposed for consistent implementation; similarity equations are evaluated only for $z_a>z_0$, where $z_0$ is aerodynamic roughness length.

\ac{most} gives the Obukhov length and stability coordinate \citep{businger1971flux,wyngaard1971behavior}
\begin{equation}
L=-\frac{u_*^3\overline{\theta_v}}{\kappa g\overline{w'\theta_v'}},
\qquad \zeta=\frac{z_a}{L},
\label{eq:obukhov}
\end{equation}
where $L$ is the Obukhov length (m), $u_*$ is the friction velocity (m\,s$^{-1}$), $\overline{\theta_v}$ is the Reynolds-averaged virtual potential temperature (K), $\kappa\simeq0.40$ is the dimensionless von Karman constant, and $g$ is gravitational acceleration (m\,s$^{-2}$). Here, $w'=w-\overline{w}$ is the turbulent fluctuation of vertical velocity (m\,s$^{-1}$), $\theta_v'=\theta_v-\overline{\theta_v}$ is the turbulent fluctuation of virtual potential temperature (K), and the overbar denotes a Reynolds average. Consequently, $\overline{w'\theta_v'}$ is the kinematic turbulent virtual-potential-temperature flux (K\,m\,s$^{-1}$), taken as positive for upward transport. The aerodynamic height is $z_a=z-d$ (m), and $\zeta=z_a/L$ is the dimensionless stability coordinate; $\zeta<0$, $\zeta\approx0$, and $\zeta>0$ correspond to unstable, near-neutral, and stable stratification, respectively.

The temperature scale and temperature structure parameter are \citep{wyngaard1971behavior,andreas1988estimating}
\begin{equation}
\theta_*=-\frac{\overline{w'\theta'}}{u_*}=-\frac{H_0}{\rho c_pu_*},
\qquad
\Cth=\theta_*^2z_a^{-2/3}f_T(\zeta),
\label{eq:mostct}
\end{equation}
where $H_0$ is upward-positive sensible heat flux, $\rho$ is air density, and $c_p$ is specific heat at constant pressure. \ac{most} provides $\Cth$, whereas Eq.~\eqref{eq:dryconvert} requires $\CT$. The function in Eq.~\eqref{eq:mostct} is defined and discussed in Section~\ref{sim_reg}. From the standard definition of potential temperature, $T=\Pi\theta$, where $\Pi=(p/p_0)^{R_d/c_p}$ is the Exner function \citep{stull1988introduction}. If pressure, and therefore $\Pi$, is treated as locally constant over the inertial-range separation used to define the structure function, then $\delta T\simeq\Pi\,\delta\theta$. Squaring this relation gives the leading local conversion
\begin{equation}
\CT\simeq\Pi^2\Cth,
\qquad
\Pi=\left(\frac{p}{p_0}\right)^{R_d/c_p},
\label{eq:exnerconvert}
\end{equation}
where $p_0=\SI{1000}{hPa}$ and $R_d$ is the dry-air gas constant. This factor is close to unity near sea level but is retained because neglecting it increasingly biases elevated-site calculations.

Eq.~\eqref{eq:mostct} reveals two distinct meteorological controls. The magnitude $\theta_*^2$ sets the scalar-variance amplitude through the ratio of sensible heat flux to friction velocity, while $f_T(\zeta)$ changes the vertical structure according to stability. Thus two periods with the same mean temperature and wind speed can yield different $\Cn$ if their surface fluxes differ, and the same heat flux can produce different $\Cn$ when surface roughness or wind modifies $u_*$. This dependence is the principal physical distinction from the fixed coefficient $A$ in \ac{hv}.

\subsubsection{Unstable, neutral, and stable similarity regimes} \label{sim_reg}

\begin{figure}[t]
\centering
\includegraphics[width=\columnwidth]{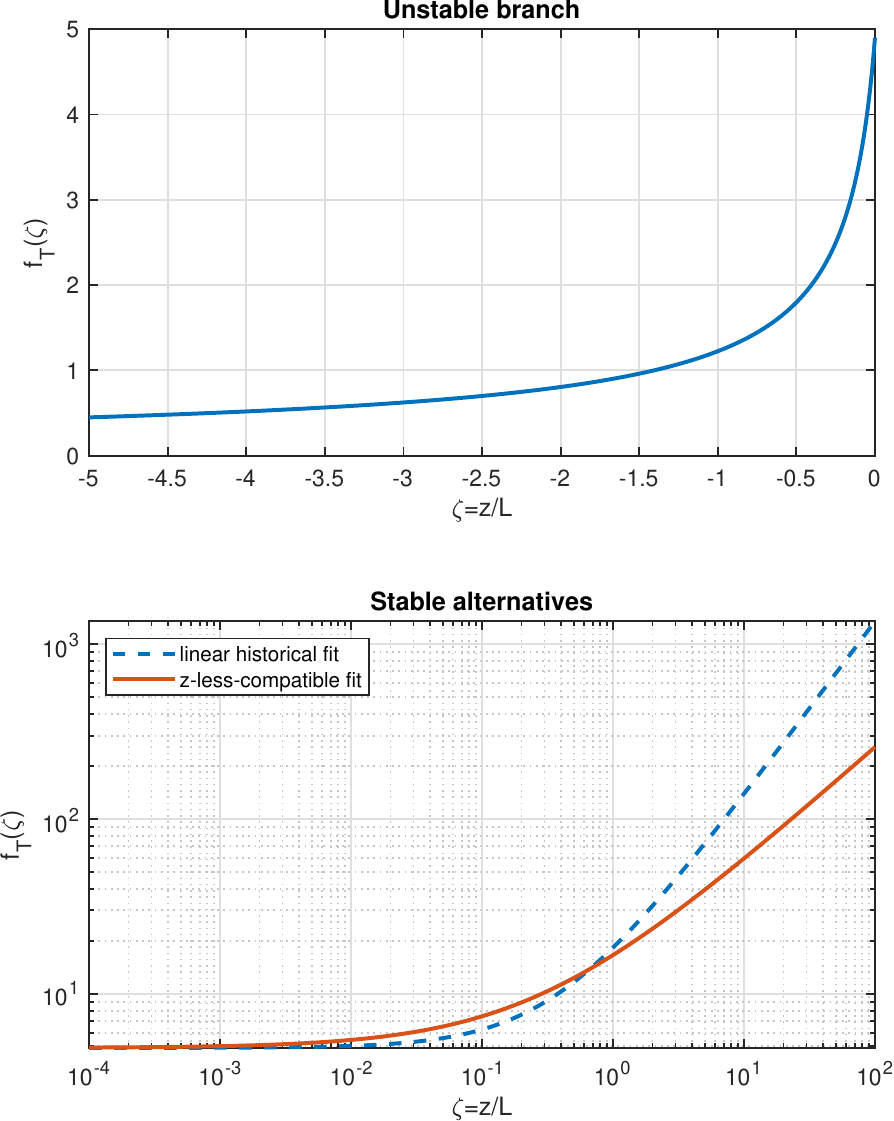}
\caption{Temperature-structure similarity functions for unstable and stable stratification.}
\label{fig:similarity}
\end{figure}

Neutral conditions correspond to $|z_a/L|\ll1$, for which the similarity function approaches $4.9$ and Eq.~\eqref{eq:mostct} retains the $z_a^{-2/3}$ dependence. For unstable and stable conditions, respectively, we use the similarity functions given by Friehe \citep[Eq.~(3) and the following definition of $g_3$]{friehe1977estimation} and Muschinski et al. \citep[Eq.~(23)]{muschinski2024anisotropy}:
\begin{equation}
f_T(\zeta)=
\begin{cases}
4.9(1-7\zeta)^{-2/3},&\zeta\le0,\\
4.9(1+2.4\zeta^{2/3}),&\zeta>0.
\end{cases}
\label{eq:ft}
\end{equation}
The stable expression approaches the \ac{sl} $z$-less limit and avoids indefinite $z^{1/3}$ growth produced by extrapolating the alternative linear fit to arbitrarily large $\zeta$ \citep{muschinski2024anisotropy}. Fig.~\ref{fig:similarity} displays both stable alternatives for transparency. It can be seen that the unstable and stable functions share the same neutral value but have different asymptotic behavior. The adopted stable function grows as $\zeta^{2/3}$ in the strongly stable limit, so its product with $z_a^{-2/3}$ tends toward a height-independent, $z$-less scaling when the other state variables are fixed.

Combining Eqs.~\eqref{eq:dryconvert}, \eqref{eq:mostct}, \eqref{eq:exnerconvert}, and \eqref{eq:ft} gives the dry-mode replacement for the \ac{hv} ground exponential:
\begin{equation}
C_{n,\rm SL}^{2}(z)=K_n(p,T,\lambda)\,\Pi^2\theta_*^2z_a^{-2/3}f_T(z_a/L).
\label{eq:surface}
\end{equation}
The component equations are published relations \citep{wyngaard1971behavior,andreas1988estimating,muschinski2024anisotropy}; their combination as the mandatory surface branch of the present hierarchy is proposed here.

When $u_*$ is not measured, the stability-corrected logarithmic wind relation is used \citep{businger1971flux,andreas1988estimating}:
\begin{equation}
u_*=\frac{\kappa U(z_r)}{\ln[(z_r-d)/z_0]-\psi_m[(z_r-d)/L]+\psi_m(z_0/L)},
\label{eq:ustar}
\end{equation}
where $U(z_r)$ is mean wind at reference height $z_r$ and $\psi_m$ is the integrated momentum correction. The unstable Businger--Dyer expression and the stable approximation $\psi_m\simeq-5\zeta$ are used in the accompanying code \citep{businger1971flux,andreas1988estimating}.

\subsection{Atmospheric boundary layer}

Above the \ac{sl}, the relationship between mean gradients and turbulent scalar variance depends strongly on \ac{bl} regime. Daytime convection produces a mixed layer capped by an entrainment zone, whereas stable and near-neutral \acp{bl} are commonly governed by shear, radiative cooling, low-level jets, and intermittent turbulent events. Treating both cases with one gradient formula would conceal these distinct mechanisms.

\subsubsection{Strongly convective regime}

Let $h_i$ be \ac{abl} height and let $h_s$ be the operational \ac{sl} top. The numerical rule
\begin{equation}
h_s=\min\!\left[\max(0.1h_i,\SI{10}{m}),\SI{100}{m},0.8h_i\right]
\label{eq:hs}
\end{equation}
is proposed here as a transparent initial choice. The additional $0.8h_i$ constraint guarantees a nonempty layer between the operational \ac{sl} top and the diagnosed \ac{abl} top for shallow \acp{bl}. Instead of imposing a discontinuous regime switch, define $\Xi=-{h_i}/{L}$
\begin{equation}
G_{\Xi}=\left[1+\exp\!\left(-\frac{\Xi-\Xi_0}{\Delta_\Xi}\right)\right]^{-1},
\label{eq:xi1}
\end{equation}

\begin{equation}
G_H=\left[1+\exp\!\left(-\frac{H_0}{\Delta_H}\right)\right]^{-1},
\label{eq:xi2}
\end{equation}
where $G_\Xi$ and $G_H$ aim to provide a continuous transition between the strongly convective and stable/near-neutral \ac{abl} formulations and then the \ac{ri} function is
\begin{equation}
C_{n,\rm ABL}^{2}=G_{\rm conv}C_{n,\rm CBL}^{2}+
(1-G_{\rm conv})C_{n,\rm diss}^{2},
\label{eq:regimegate}
\end{equation}
where $G_{\rm conv}=G_\Xi G_H$ is a continuous convective-membership factor. The initial values $\Xi_0=10$, $\Delta_\Xi=2$, and $\Delta_H=\SI{20}{W.m^{-2}}$ are proposed classification parameters requiring calibration.

The quantity $\Xi$ compares \ac{bl} depth with the magnitude of the Obukhov length and therefore measures whether surface buoyancy forcing is strong enough to influence the full layer. The heat-flux gate suppresses convective membership when the surface flux is downward or weak. 

Good et al. summarize a Walters--Kunkel/Kaimal-family three-zone convective profile \citep{good1988atmospheric}. Written as a dimensionless shape and normalized at $z_r=h_s$, it is
\begin{equation}
F_{\rm CBL}(z)=
\begin{cases}
(z/z_r)^{-4/3},&z_r<z\le0.5h_i,\\
(0.5h_i/z_r)^{-4/3},&0.5h_i<z\le0.7h_i,\\
2.9(0.5h_i/z_r)^{-4/3}(z/h_i)^3,&0.7h_i<z\le h_i.
\end{cases}
\label{eq:cblshape}
\end{equation}
Then, the \ac{ri} structure parameter for \ac{cbl} can be written by
\begin{equation}
C_{n,\rm CBL}^{2}(z)=C_{n,\rm SL}^{2}(h_s)F_{\rm CBL}(z).
\label{eq:cblmatch}
\end{equation}
Eq.~\eqref{eq:cblshape} is literature based while Eq.~\eqref{eq:cblmatch} is a new continuity construction that must be calibrated against mixed-layer and entrainment-zone observations.

The first branch of Eq.~\eqref{eq:cblshape} retains the free-convective $z^{-4/3}$ decrease. The middle branch approximates the weak vertical variation expected in a mixed-layer interior, and the upper branch represents enhancement toward the entrainment zone. The factor $2.9$ makes the latter branch nearly continuous at $0.7h_i$ because $2.9(0.7)^3\simeq0.995$. Matching the amplitude at $h_s$ avoids an arbitrary second normalization constant, but it also transfers \ac{sl} uncertainty into the convective profile and must therefore be tested empirically.

\subsubsection{Stable and near-neutral regimes}

As $H_0$ becomes non-positive or $\Xi$ falls below $\Xi_0$, $G_{\rm conv}$ approaches zero smoothly. Near-neutral conditions have weak buoyancy production and may be dominated by mechanically generated turbulence. Stable conditions may contain intermittent turbulence, low-level jets, and thin layers that are not described by a single universal full-depth \ac{abl} profile \citep{muschinski2024anisotropy}. In the complete-data mode, the present hierarchy therefore uses the resolved dissipation closure of Eq.~\eqref{eq:diss} for the stable and near-neutral \ac{abl}. In the present formulation, the \ac{abl} and free-atmosphere components use the same local dissipation-based closure when the required turbulence variables are available. At this stage, their separation therefore represents the physical organization of the atmosphere rather than two different mathematical parameterizations. A distinct stable-\ac{abl} component may be incorporated when a suitable formulation has been independently calibrated and validated.

We adopt different representations for strongly convective and stable or near-neutral conditions because these regimes exhibit fundamentally different turbulence structures. For strongly convective conditions, we use an explicit vertical profile, since a local gradient-based closure cannot adequately describe the weak mean gradients of the well-mixed layer or the enhanced turbulence associated with the entrainment region. For stable and near-neutral conditions, we use local turbulence prognostic variables whenever they are available. Stable \acp{bl} are strongly affected by intermittency, wind shear, low-level jets, gravity waves, and thin turbulent layers. 

\subsection{Free atmosphere}

The free atmosphere begins above the diagnosed \ac{bl} top and is not directly coupled to local surface fluxes on the averaging time scales considered here. Optical turbulence may nevertheless be concentrated in thin layers associated with vertical shear, static-stability changes, gravity-wave activity, and turbulent mixing near the tropopause. A smooth altitude-only background cannot move these layers in response to the atmospheric state. The preferred formulation therefore uses local thermodynamic and turbulence variables rather than prescribing their altitude a priori.

\subsubsection{Dissipation closure}

For fully developed, locally homogeneous and isotropic scalar turbulence, the Obukhov--Corrsin relation is \citep{muschinski2015temperature,muschinski2025predicting}
\begin{equation}
\CT=\gamma\varepsilon^{-1/3}\chi,
\qquad \gamma\simeq1.6,
\label{eq:oc}
\end{equation}
where $\varepsilon$ is turbulent kinetic-energy dissipation rate in \si{m^2.s^{-3}}, $\chi$ is temperature-variance dissipation rate in \si{K^2.s^{-1}}, and $\gamma$ is the Obukhov--Corrsin constant. Combining Eqs.~\eqref{eq:dryconvert} and \eqref{eq:oc} gives \citep[Eq.~(5)]{muschinski2025predicting}
\begin{equation}
C_{n,\rm diss}^{2}(H)=K_n[p(H),T(H),\lambda]\,\gamma\varepsilon(H)^{-1/3}\chi(H).
\label{eq:diss}
\end{equation}
Eq.~\eqref{eq:diss} is the preferred branch above the \ac{sl} when $\varepsilon$ and $\chi$ are supplied by observations or a turbulence model. Its assumptions can fail where scale separation is inadequate or turbulence is strongly anisotropic \citep{muschinski2024anisotropy,muschinski2025predicting}.

We use the dissipation-based relation because it connects $C_T^2$ directly to the scalar-variance cascade rather than prescribing its variation through an empirically fitted altitude profile. The resulting estimate, however, depends on the resolution and turbulence parameterization of the atmospheric dataset from which $\varepsilon$ and $\chi$ are obtained. In particular, their calculated values are influenced by the adopted turbulence closure, vertical grid spacing, and spatial and temporal averaging procedures. These modeling characteristics should therefore be reported together with every derived $C_n^2$ profile. For example, the approximately \SI{500}{m} vertical spacing considered by Muschinski et al. may reproduce the broad climatological distribution of optical turbulence but cannot resolve turbulent sheets with vertical scales of only a few metres \citep{muschinski2025predicting}.

\subsubsection{Gradient/outer-scale alternative and convention warning}

Tatarskii's first outer-scale definition equates mean-gradient and turbulent structure functions \citep{tatarskii1961wave,muschinski2024anisotropy}:
\begin{equation}
L_{0,1}=\left[\frac{\Cth}{(\mathrm{d}\overline{\theta}/\mathrm{d}z)^2}\right]^{3/4}.
\label{eq:L01}
\end{equation}
With this definition, $\Cth=(\mathrm{d}\overline{\theta}/\mathrm{d}z)^2L_{0,1}^{4/3}$ follows directly. Tatarskii later introduced a second, inconsistent outer-scale convention under the same symbol \citep{muschinski2024anisotropy}. Consequently, the historical \ac{afgl} form \citep[Eq.~(24)]{good1988atmospheric}
\begin{equation}
C_{n,\rm AFGL}^{2}=2.8M^2L_{0,2}^{4/3}
\label{eq:afglgeneric}
\end{equation}
must be used only with its associated $L_{0,2}$ convention and calibrated shear regression \citep{good1988atmospheric,beland1993propagation}. Eqs.~\eqref{eq:L01} and \eqref{eq:afglgeneric} must not be combined by treating $L_{0,1}$ and $L_{0,2}$ as identical. The present figures therefore use Eq.~\eqref{eq:diss}, not Eq.~\eqref{eq:afglgeneric}, as the upper-air branch.

\section{Proposed model}

The proposed model is a hierarchy of conditional estimates rather than a single closed-form altitude curve. This distinction is important. We are not fitting a new universal function of height; we are asking which physical description is most defensible at each height and then joining the descriptions without creating an artificial discontinuity. The construction is guided by four requirements: (i) the surface estimate must retain the measured or diagnosed dependence on flux and stability; (ii) the convective mixed layer must not collapse to zero when its mean potential-temperature gradient is small; (iii) the upper-air estimate must be able to follow meteorological layers; and (iv) the result must remain positive and continuous when the active description changes.

For a given meteorological profile, we first calculate three candidate fields: $C_{n,\rm SL}^{2}$ from Eq.~\eqref{eq:surface}, $C_{n,\rm ABL}^{2}$ from Eq.~\eqref{eq:regimegate}, and $C_{n,\rm FA}^{2}$ from Eq.~\eqref{eq:diss}. These candidates are not added as independent sources of variance. Only one should dominate in the interior of each region. The weights described below are therefore membership functions: they express how strongly a height belongs to the \ac{sl}, remaining \ac{bl}, or free-atmospheric description.

Because logistic functions have finite tails, each candidate must be defined throughout a narrow overlap interval. In the numerical implementation, the surface candidate is evaluated only for $z_a>z_0$ and is held at its nearest boundary value outside $z_0<z\le h_s$; the convective shape is similarly evaluated with its height argument clipped to $h_s\le z\le h_i$. These continuations are numerical boundary rules, not claims that \ac{most} or the three-zone convective law remains physically valid outside its stated domain.

\subsection{Choice of regional boundaries}

The two transition heights are $h_s$, the operational top of the \ac{sl}, and $h_i$, the diagnosed \ac{bl} height. The choice $h_s=0.1h_i$ reflects the conventional view that the constant-flux \ac{sl} occupies approximately the lowest tenth of the \ac{bl}, but the exact fraction varies with stability, heterogeneity, and time dependence. We therefore clip it between \SI{10}{m} and \SI{100}{m} in Eq.~\eqref{eq:hs}. The lower limit avoids applying a transition below typical tower or model resolution; the upper limit prevents surface similarity from being extrapolated hundreds of metres through a deep convective layer. 

\begin{figure}[t]
\centering
\resizebox{\columnwidth}{!}{%
\begin{tikzpicture}[font=\sffamily\small,>=Latex]
  \fill[orange!32] (0.8,0.45) rectangle (3.35,2.05);
  \fill[cyan!24]   (0.8,2.05) rectangle (3.35,5.00);
  \fill[blue!10]   (0.8,5.00) rectangle (3.35,7.05);

  \fill[orange!60,opacity=0.42] (0.8,1.72) rectangle (3.35,2.38);
  \fill[blue!55,opacity=0.35]   (0.8,4.48) rectangle (3.35,5.52);

  \draw[thick] (0.8,0.45) rectangle (3.35,7.05);
  \draw[very thick,brown!70!black] (0.55,0.45) -- (3.60,0.45);
  \node[below] at (2.08,0.45) {local ground, $z=0$};

  \draw[dashed,thick] (0.45,2.05) -- (3.60,2.05);
  \draw[dashed,thick] (0.45,5.00) -- (3.60,5.00);
  \node[anchor=east,fill=white,inner sep=1pt] at (0.70,2.05) {$h_s$};
  \node[anchor=east,fill=white,inner sep=1pt] at (0.70,5.00) {$h_i$};

  \node[align=center] at (2.08,1.15)
    {surface layer\\$C_{n,\mathrm{SL}}^2$};
  \node[align=center] at (2.08,3.55)
    {ABL above\\surface layer\\$C_{n,\mathrm{ABL}}^2$};
  \node[align=center] at (2.08,6.12)
    {free atmosphere\\$C_{n,\mathrm{FA}}^2$};

  \draw[<->,thick] (3.68,1.72) -- (3.68,2.38);
  \node[anchor=west,align=left] at (3.82,2.05)
    {$D_{s,10--90}=4.394\Delta_s$\\$B_s(h_s)=1/2$};

  \draw[<->,thick] (3.68,4.48) -- (3.68,5.52);
  \node[anchor=west,align=left] at (3.82,5.00)
    {$D_{i,10--90}=4.394\Delta_i$\\$B_i(h_i)=1/2$};

  \draw[->,thick] (0.08,0.45) -- (0.08,7.15)
    node[above] {$z$};

  \node[draw,rounded corners=1.5pt,fill=white,align=left,
        inner sep=4pt,anchor=south west] at (4.05,2.70)
    {$W_{\rm SL}=B_s$\\
     $W_{\rm ABL}=(1-B_s)B_i$\\
     $W_{\rm FA}=(1-B_s)(1-B_i)$};
\end{tikzpicture}%
}
\caption{Regional model, transition midpoints, and 10--90\% blending intervals.}
\label{fig:regional_layers}
\end{figure}
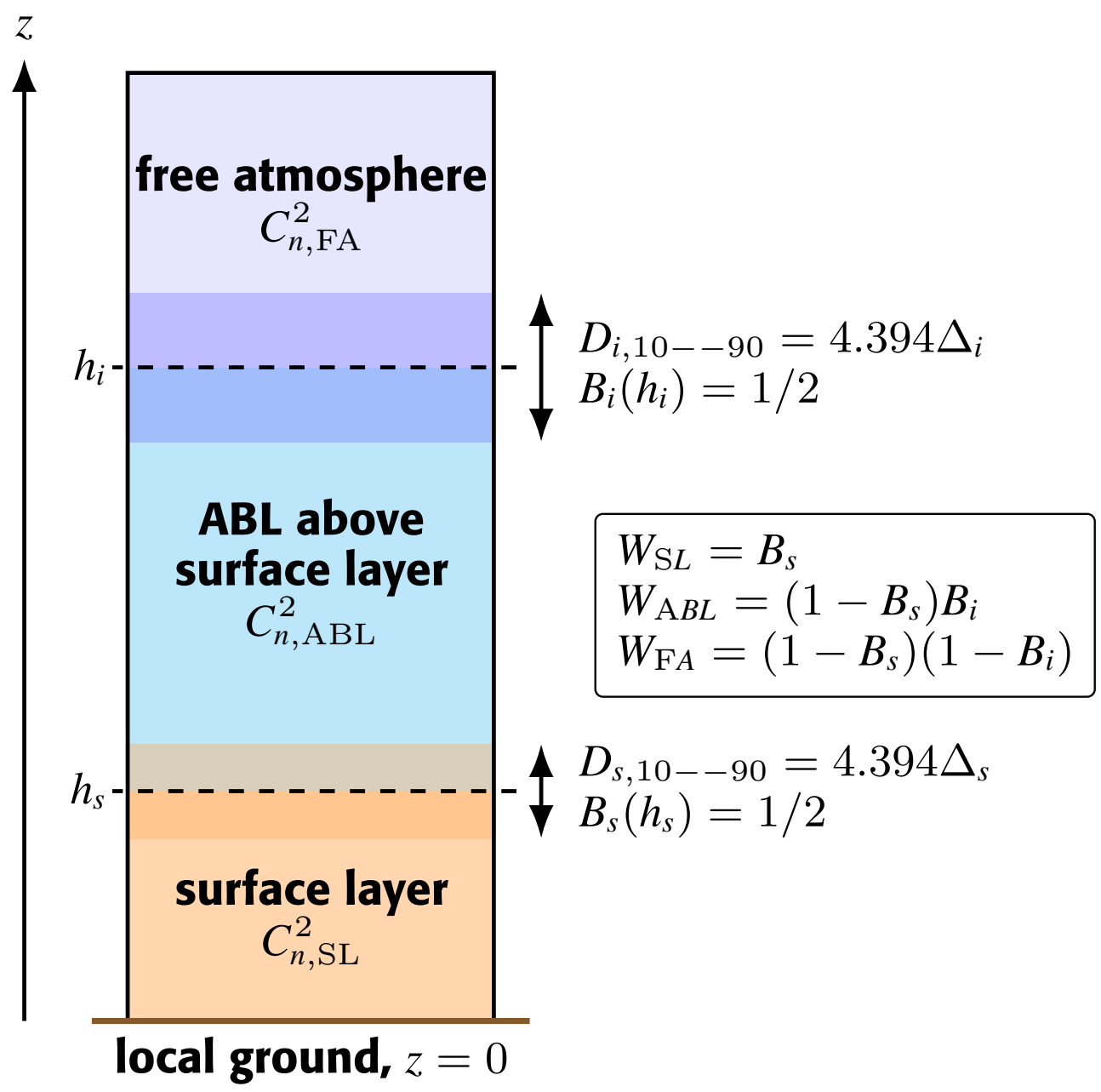

The height $h_i$ is not inferred from $\Cn$ in the present formulation. It must be supplied by an independent \ac{bl} diagnosis, for example a potential-temperature inversion, a bulk Richardson-number method, a ceilometer product, or the \ac{bl} height produced by a numerical weather model. This separation prevents circular reasoning in which a feature of the predicted $\Cn$ profile is used both to define and to test the \ac{abl} top. Because different $h_i$ algorithms need not agree, the algorithm and temporal averaging interval are part of the required model metadata. Fig.~\ref{fig:regional_layers} summarizes the regional definitions and the finite transition intervals used by the composite formulation.

\subsection{Continuous composite formulation}

To combine the regional estimates without introducing discontinuities at $h_s$ and $h_i$, we define two height-dependent blending functions:
\begin{equation}
B_s(z)=\left[1+\exp\!\left(\frac{z-h_s}{\Delta_s}\right)\right]^{-1},
\label{eq:Bs}
\end{equation}
and
\begin{equation}
B_i(z)=\left[1+\exp\!\left(\frac{z-h_i}{\Delta_i}\right)\right]^{-1}.
\label{eq:Bi}
\end{equation}
The function $B_s$ controls the transition from the \ac{sl} description, whereas $B_i$ controls the transition from the \ac{abl} description to the free atmosphere. Both functions decrease monotonically from one to zero and satisfy $B_s(h_s)=B_i(h_i)=1/2$. The parameters $\Delta_s$ and $\Delta_i$ control the steepness of the corresponding transitions rather than their full physical depths. We selected this form because it is bounded, monotonic, continuously differentiable, and specified by a directly interpretable midpoint and transition scale. 

Combining regional estimates, the proposed \ac{ri} structure parameter becomes
\begin{equation}
\begin{split}
\Cn(z)={}&B_sC_{n,\rm SL}^{2}\\
&+(1-B_s)\left[
B_iC_{n,\rm ABL}^{2}
+(1-B_i)C_{n,\rm FA}^{2}\right].
\end{split}
\label{eq:composite}
\end{equation}
This construction is sequential. The factor $B_s$ first assigns the \ac{sl} contribution. The remaining fraction, $1-B_s$, is then divided by $B_i$ between the \ac{abl} and free-atmosphere candidates. Equivalently, define
\begin{equation}
\begin{split}
W_{\rm SL}&=B_s,\qquad
W_{\rm ABL}=(1-B_s)B_i,\\
W_{\rm FA}&=(1-B_s)(1-B_i).
\end{split}
\label{eq:regionalweights}
\end{equation}
These weights are nonnegative and satisfy
\begin{equation}
W_{\rm SL}+W_{\rm ABL}+W_{\rm FA}
=B_s+(1-B_s)\left[B_i+(1-B_i)\right]=1.
\label{eq:weightsum}
\end{equation}
Eq.~\eqref{eq:composite} is therefore a convex combination of the three positive regional estimates. The normalization preserves positivity and prevents double counting within the transition intervals.

The initial numerical widths are proposed as
\begin{equation}
\Delta_s=\max(0.02h_i,\SI{5}{m}),\qquad
\Delta_i=\max(0.05h_i,\SI{20}{m}).
\label{eq:widths}
\end{equation}
It should be noted that the parameter $\Delta$ is not the full transition depth. For a logistic function the distance between weights 0.9 and 0.1 is
\begin{equation}
D_{10--90}=2\ln(9)\Delta\simeq4.394\Delta.
\label{eq:transitiondepth}
\end{equation}
Consequently, when the fractional terms in Eq.~\eqref{eq:widths} control, the surface transition spans approximately $0.088h_i$ and the \ac{abl}-top transition approximately $0.22h_i$. We chose the surface transition to be narrower because \ac{most} loses validity over a comparatively short interval above the constant-flux layer. The broader \ac{abl}-top transition acknowledges that entrainment occupies a finite fraction of the layer and that diagnosed $h_i$ is uncertain. The absolute floors of \SI{5}{m} and \SI{20}{m} prevent an unrealistically sharp switch for shallow layers or coarse input profiles. These values are starting assumptions, not measured universal constants. If the observations show a sharp physical interface at a particular site, the inferred $\Delta_i$ should decrease; if the diagnosed \ac{abl} top varies substantially within the averaging interval, a larger effective width may be appropriate.

The limiting behavior provides a direct consistency check. Well below $h_s$, $B_s\rightarrow1$ and Eq.~\eqref{eq:composite} reduces to the surface solution. For $h_s\ll z\ll h_i$, $B_s\rightarrow0$ and $B_i\rightarrow1$, leaving the \ac{abl} solution. Well above $h_i$, both logistic functions approach zero and the free-atmospheric solution is recovered. At $z=h_s$, $B_s=1/2$; at $z=h_i$, $B_i=1/2$. The value is continuous everywhere. Its derivative is also finite, although it need not match the derivative of either candidate because the candidates can have different slopes. Table~\ref{tab:proposedparameters} collects the initial regional-classification and blending parameters used in the illustrative calculations. These values are calibration starting points rather than universal atmospheric constants. 

\begin{table}[t]
\centering
\caption{Proposed regional and blending parameters.}
\label{tab:proposedparameters}

\scriptsize
\setlength{\tabcolsep}{2.5pt}
\renewcommand{\arraystretch}{1.10}

\begin{tabularx}{\columnwidth}{
  @{}
  >{\raggedright\arraybackslash}p{0.19\columnwidth}
  >{\centering\arraybackslash}p{0.17\columnwidth}
  >{\raggedright\arraybackslash}X
  @{}
}
\toprule
Parameter & Value & Role \\
\midrule

$h_s/h_i$
& $0.10$
& Nominal \ac{sl} fraction; constrained by
Eq.~\eqref{eq:hs} and $h_s\leq0.8h_i$. \\

$h_{s,\min}$,
$h_{s,\max}$
& $10$, $100$ m
& Lower and upper limits for application of \ac{sl} similarity. \\

$\Xi_0$
& $10$
& Center of the convective-regime transition. \\

$\Delta_\Xi$
& $2$
& Width of the convective-regime transition. \\

$\Delta_H$
& $20$ W\,m$^{-2}$
& Width of the sensible-heat-flux gate. \\

$\Delta_s/h_i$
& $0.02$
& Surface blending scale; 10--90\% depth is $0.088h_i$. \\

$\Delta_i/h_i$
& $0.05$
& \ac{bl}-top blending scale; 10--90\% depth is $0.22h_i$. \\

$\Delta_{s,\min}$,
$\Delta_{i,\min}$
& $5$, $20$ m
& Minimum transition scales used as resolution safeguards. \\

\bottomrule
\end{tabularx}
\end{table}

\section{Results and discussion}

For regime $r\in\{\mathrm{convective},\mathrm{neutral},\mathrm{stable}\}$, define the prescribed input set
\begin{equation}
\begin{aligned}
\mathcal{M}^{(r)}=\bigl\{&
P(z),T(z),V(z),\varepsilon(z),\chi(z);\\
&
H_0^{(r)},U_{10}^{(r)},h_i,z_0,d
\bigr\}.
\end{aligned}
\label{eq:inputset}
\end{equation}
The semicolon separates height-dependent profiles from surface and layer parameters. The proposed model is then evaluated as the deterministic mapping
\begin{equation}
\begin{aligned}
&C_{n,\rm prop}^{2(r)}(z)
=\mathcal{F}\!\left[\mathcal{M}^{(r)};\boldsymbol{\Theta}\right],\\
&\boldsymbol{\Theta}
=\bigl\{\Xi_0,\Delta_\Xi,\Delta_H,h_{s,\min},h_{s,\max},\Delta_s,\Delta_i\bigr\}.
\end{aligned}
\label{eq:deterministicmapping}
\end{equation}
where $\mathcal{F}$ denotes Eqs.~\eqref{eq:surface}, \eqref{eq:cblmatch}, \eqref{eq:diss}, and \eqref{eq:composite}, and $\boldsymbol{\Theta}$ contains the proposed classification and transition parameters. No random perturbation, data assimilation, parameter fitting, or measured $C_n^2$ is used in producing these curves.

Two types of calculation are reported. A reference-case calculation evaluates Eq.~\eqref{eq:deterministicmapping} using one complete input set from Table~\ref{tab:parameters}. A one-at-a-time sensitivity calculation replaces one input $x_k$ by a selected value $s$ while holding all other entries fixed:
\begin{equation}
C_{n,k}^{2}(z;s)=\mathcal{F}\!\left[\mathcal{M}_{-k},x_k=s;\boldsymbol{\Theta}\right].
\label{eq:oat}
\end{equation}
Thus, a separation between curves shows mathematical sensitivity to the varied input under the stated assumptions. Results are obtained by MATLAB software package.

\begin{table}[t]
\centering
\caption{Principal numerical settings used in the figures.}
\label{tab:parameters}
\scriptsize
\setlength{\tabcolsep}{2.0pt}
\renewcommand{\arraystretch}{1.12}
\begin{tabularx}{\columnwidth}{
  @{}
  >{\raggedright\arraybackslash}p{0.18\columnwidth}
  >{\raggedright\arraybackslash}p{0.27\columnwidth}
  @{\hspace{2pt}}|@{\hspace{2pt}}
  >{\raggedright\arraybackslash}p{0.18\columnwidth}
  >{\raggedright\arraybackslash}X
  @{}
}
\toprule
Quantity & Value & Quantity & Value \\
\midrule

Wavelength, $\lambda$
& \SI{500}{nm}
& Surface $P,T$
& \SI{1013.25}{hPa};\newline \SI{293.15}{K} \\

$z_0,d,z_r$
& \SI{0.05}{m}; \SI{0}{m};\newline \SI{10}{m}
& \ac{abl} height, $h_i$
& \SI{1200}{m}, unless varied \\

Path
& $H_{\rm top}=\SI{30}{km}$;\newline vertical
& \ac{hv} 5/7
& $w=\SI{21}{m.s^{-1}}$;\newline
  $A=1.7\times10^{-14}\ \si{m^{-2/3}}$ \\

Gate parameters
& $\Xi_0=10$; $\Delta_\Xi=2$;\newline
  $\Delta_H=\SI{20}{W.m^{-2}}$
& Convective
& $H_0=\SI{250}{W.m^{-2}}$;\newline
  $U_{10}=\SI{4}{m.s^{-1}}$;\newline
  $L=\SI{-16.83}{m}$ \\

Near-neutral
& $H_0=\SI{1}{W.m^{-2}}$;\newline
  $U_{10}=\SI{9}{m.s^{-1}}$;\newline
  $L=-2.825\times10^4\ \si{m}$
& Stable
& $H_0=\SI{-10}{W.m^{-2}}$;\newline
  $U_{10}=\SI{3}{m.s^{-1}}$;\newline
  $L=\SI{72.59}{m}$ \\

\bottomrule
\end{tabularx}
\end{table}

\begin{figure}[t]
\centering
\includegraphics[width=\columnwidth]{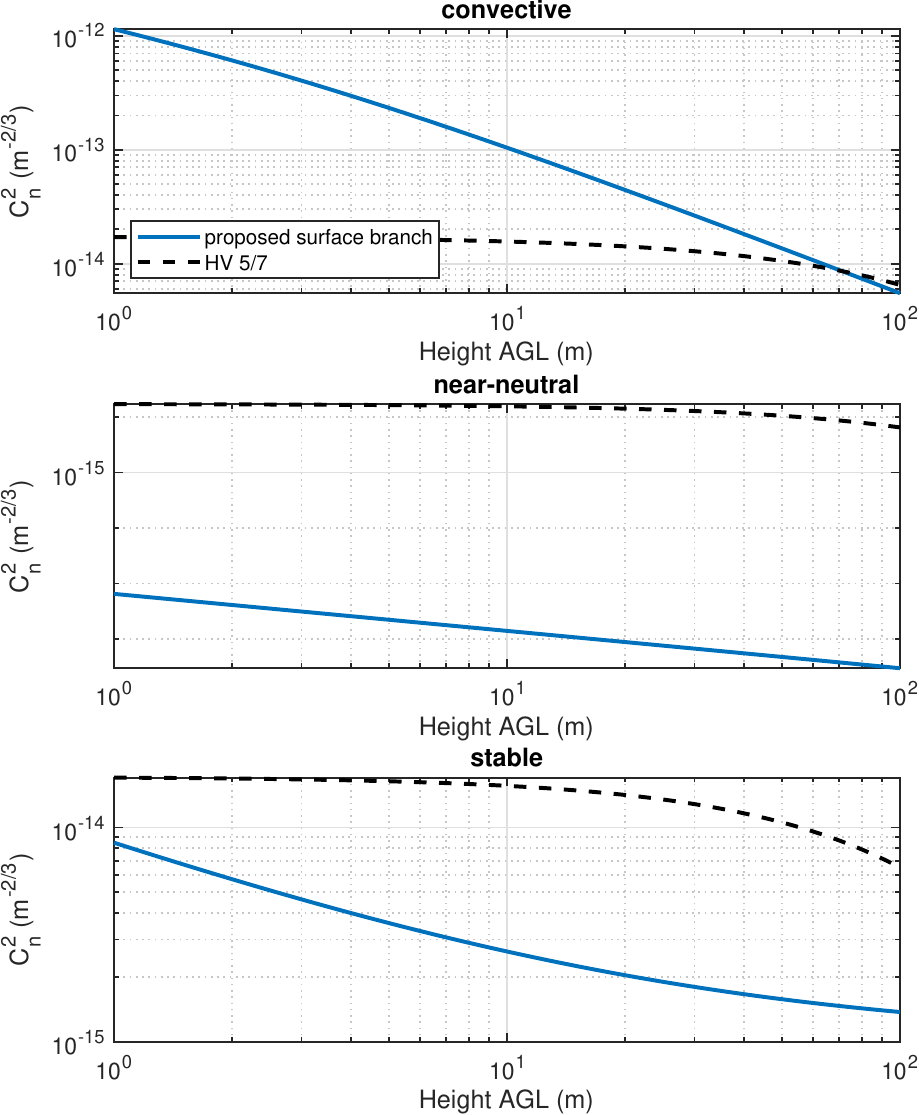}
\caption{Surface-layer $C_n^2$ for the three prescribed regimes compared with \ac{hv} 5/7.}
\label{fig:surfacehv}
\end{figure}

Fig.~\ref{fig:surfacehv} shows that the proposed surface branch responds strongly to atmospheric stability, whereas the \ac{hv} profile remains unchanged across the three prescribed regimes. This contrast follows directly from Eq.~\eqref{eq:mostct}, the surface flux sets the temperature-fluctuation amplitude through $\theta_*$, while $f_T$ modifies its vertical dependence according to stability. We obtain the largest near-ground values in the convective case because the positive sensible heat flux produces a large $|\theta_*|$. The near-neutral and stable cases remain below the convective profile for the selected flux--wind combinations. In every case, we diagnose $L$ consistently from $H_0$, $U_{10}$, and the surface state rather than prescribing it independently.

\begin{figure}[t]
\centering
\includegraphics[width=\columnwidth]{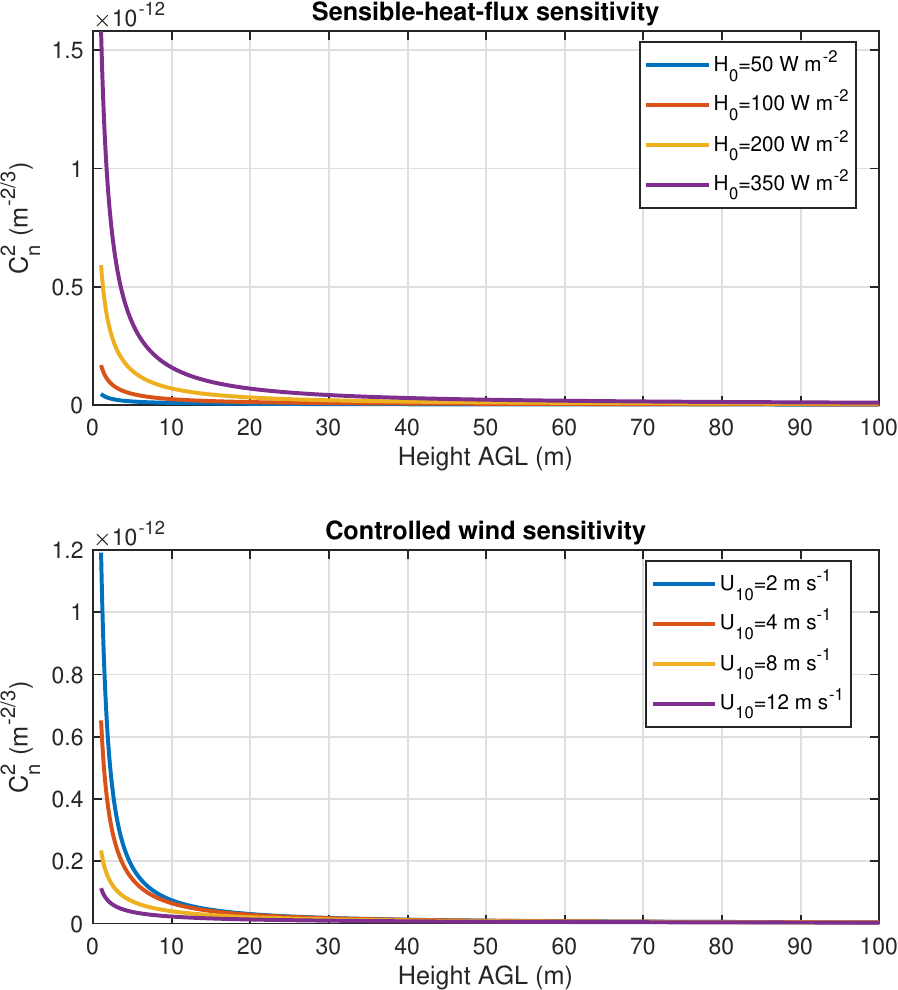}
\caption{One-at-a-time sensitivity of surface-layer $C_n^2$ to sensible heat flux and wind speed.}
\label{fig:surfacesensitivity}
\end{figure}

Fig.~\ref{fig:surfacesensitivity} separates the two principal surface controls. Increasing $H_0$ increases $|\theta_*|$ and therefore raises $C_n^2$. Wind has a competing influence: it increases the friction velocity, which tends to reduce $|\theta_*|$ for a fixed heat flux, but it also changes $L$ and hence the stability function. The resulting wind dependence is therefore conditional on the specified thermal forcing rather than universally increasing or decreasing. The magnitude of the contrast is substantial for the selected inputs. At \SI{1}{m}, the convective, near-neutral, and stable calculations are approximately $65.97$, $8.79\times10^{-4}$, and $0.489$ times the \ac{hv} 5/7 value, respectively. At \SI{100}{m}, the corresponding ratios are approximately $0.845$, $0.0404$, and $0.146$. These changes are generated by the prescribed flux--wind combinations and the resulting self-consistent stability, not by fitted agreement with \ac{hv}. In particular, the very small near-neutral surface value follows from choosing $H_0=\SI{1}{W.m^{-2}}$.

\begin{figure}[t]
\centering
\includegraphics[width=\columnwidth]{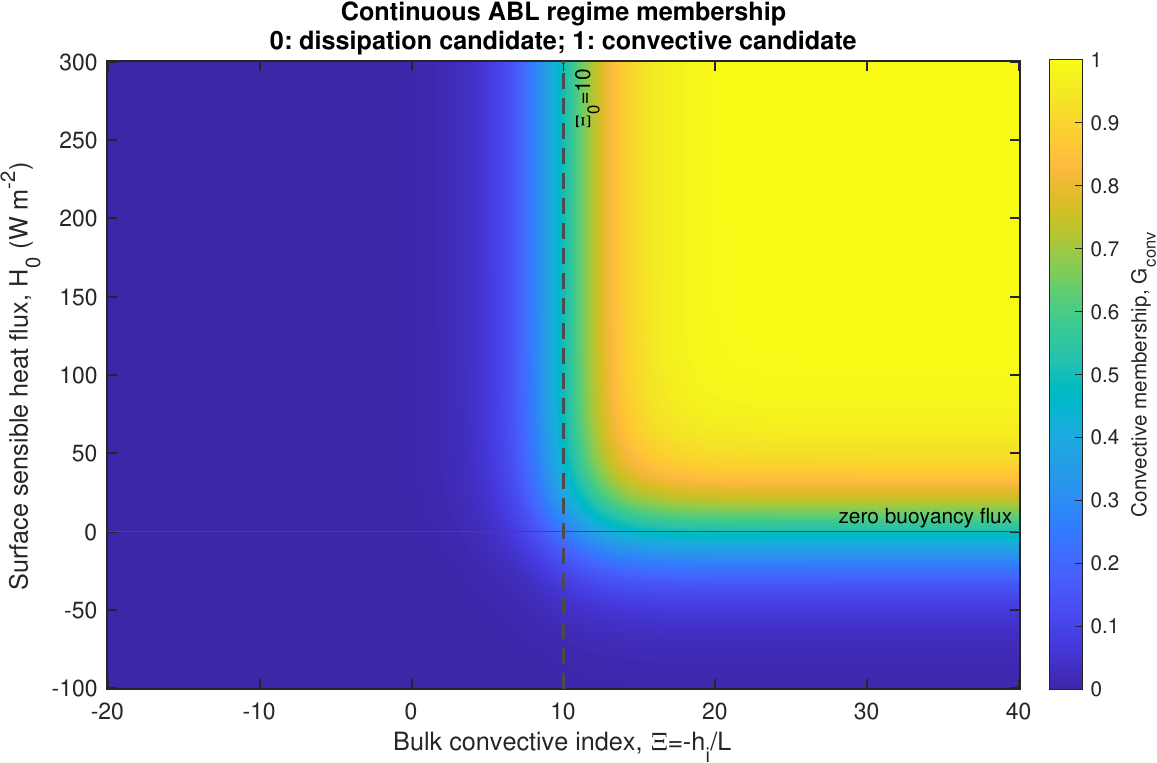}
\caption{Convective-membership factor $G_{\rm conv}$ as a function of $\Xi=-h_i/L$ and surface sensible heat flux.}
\label{fig:regimemap}
\end{figure}

Figure~\ref{fig:regimemap} confirms the intended behavior of the convective gate. The parameter $\Xi=-h_i/L$ compares the boundary-layer depth with the Obukhov length and therefore measures whether surface buoyancy forcing can influence a substantial fraction of the layer. Large convective membership occurs only when this bulk instability is accompanied by positive surface heating; neither condition alone activates the convective branch fully. The smooth intermediate values represent transitions between candidate closures rather than an additional turbulence regime.

\begin{figure}[t]
\centering
\includegraphics[width=\columnwidth]{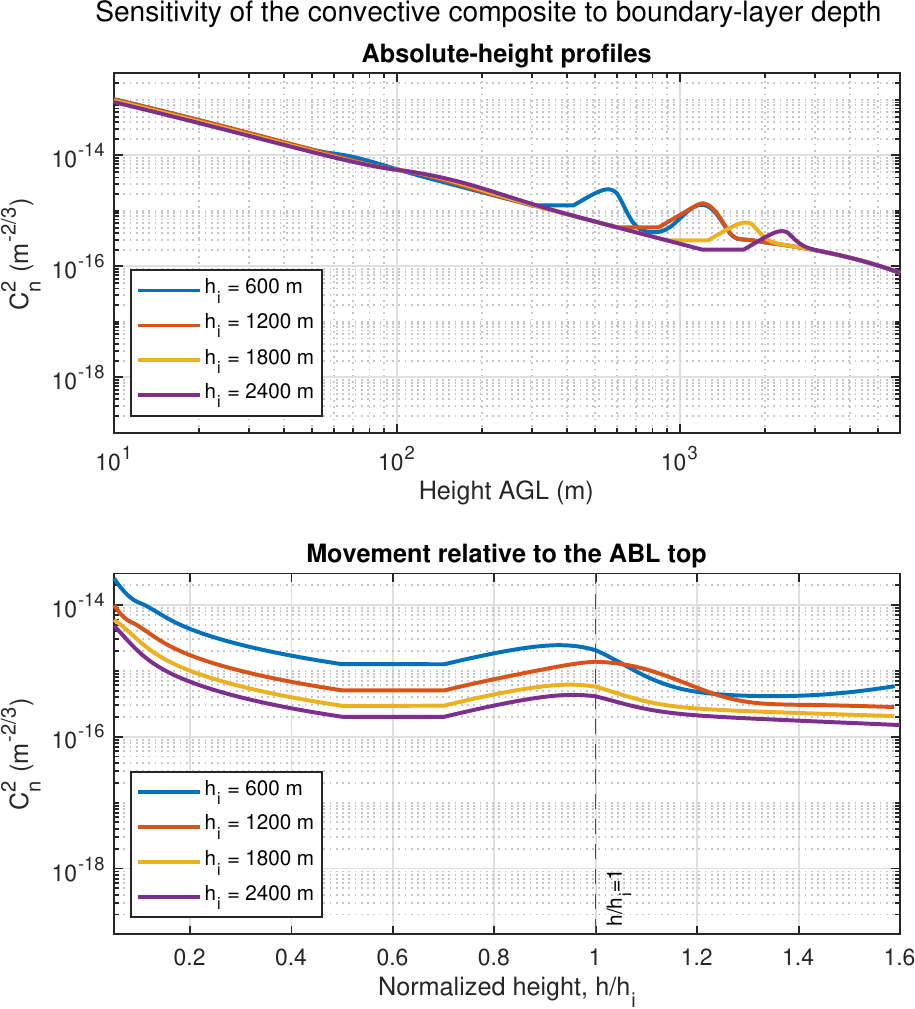}
\caption{Sensitivity of the convective profile to $h_i$ in absolute (left) and normalized (right) height coordinates.}
\label{fig:depth}
\end{figure}

\begin{figure}[t]
\centering
\includegraphics[width=\columnwidth,trim=0 0 0 18pt,clip]{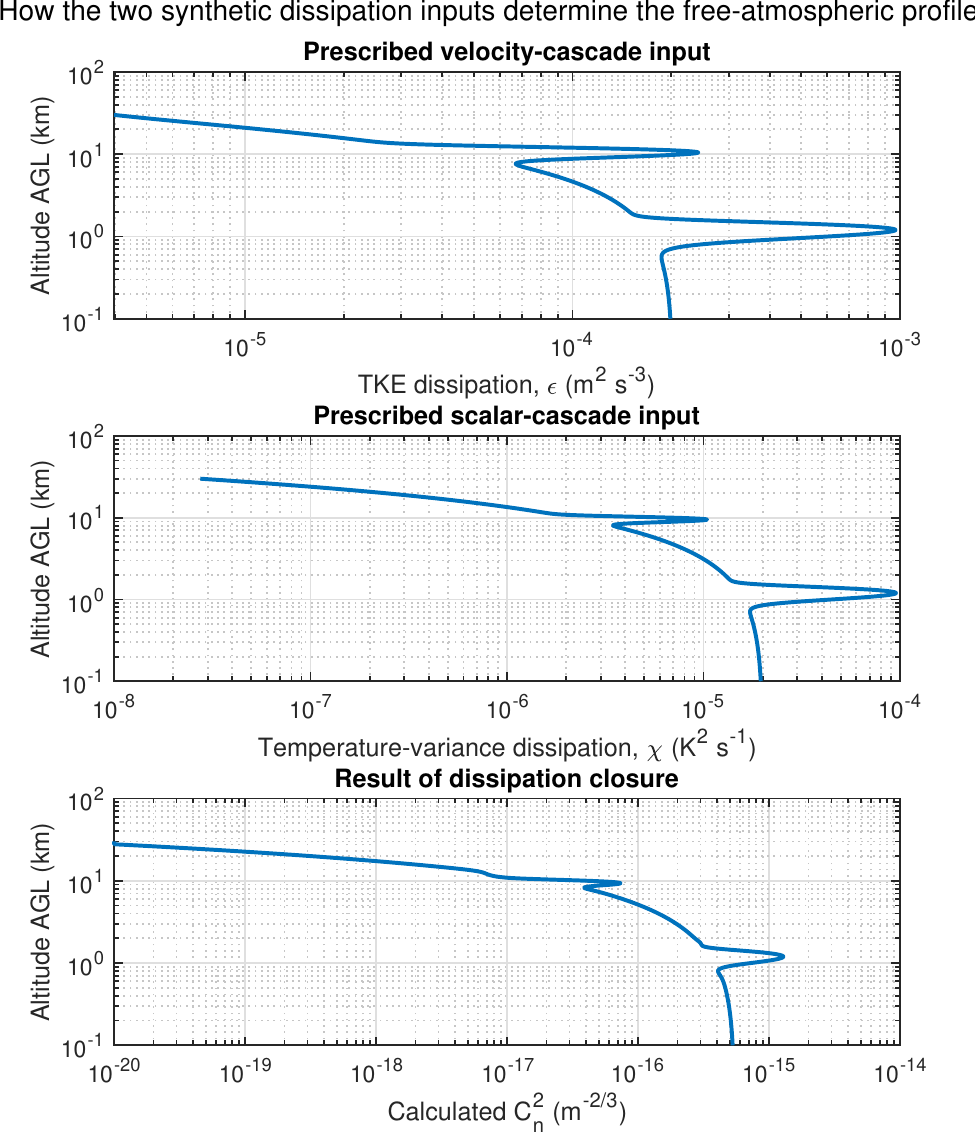}
\caption{Prescribed $\varepsilon$ and $\chi$ profiles and the resulting dissipation-based $C_n^2$.}
\label{fig:dissipation}
\end{figure}

\begin{figure}[t]
\centering
\includegraphics[width=\columnwidth]{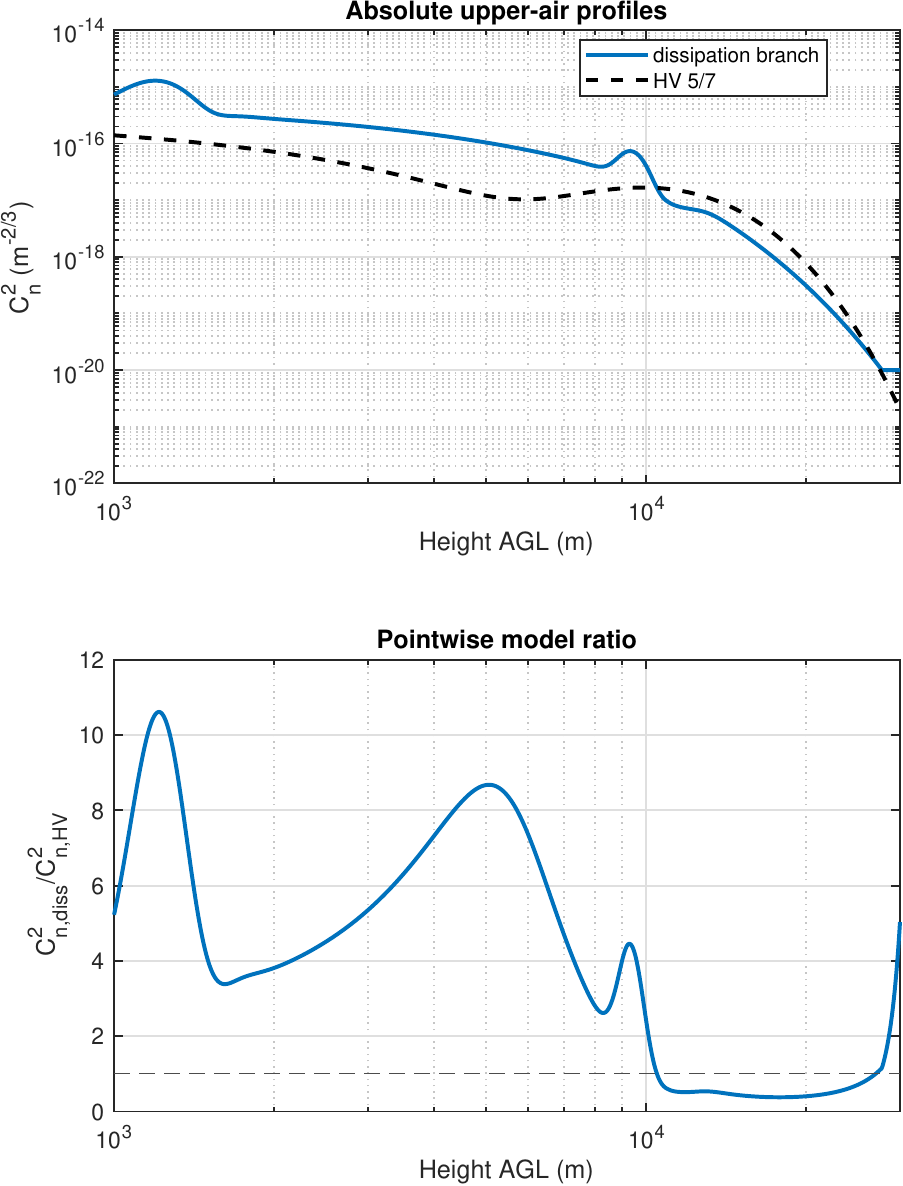}
\caption{Free-atmosphere $C_n^2$ and its pointwise ratio to \ac{hv} 5/7.}
\label{fig:freehv}
\end{figure}

Figure~\ref{fig:depth} shows that changing $h_i$ displaces the modeled mixed-layer and entrainment features in absolute height. Their closer alignment in the normalized coordinate $z/h_i$ is expected because the three-zone profile places its internal boundaries at fixed fractions of $h_i$. The calculation therefore verifies that the model follows an independently diagnosed boundary-layer depth rather than imposing a fixed altitude scale.

In Fig.~\ref{fig:dissipation}, the imposed layers in $\varepsilon$ and $\chi$ are transferred to $C_n^2$ through $C_T^2\propto\varepsilon^{-1/3}\chi$. Temperature-variance dissipation enters linearly, whereas kinetic-energy dissipation enters with the weaker inverse one-third power; consequently, localized variations in $\chi$ exert the more direct influence for otherwise comparable perturbations. Figure~\ref{fig:freehv} then shows localized departures from the smooth \ac{hv} profile at the prescribed upper-air layers. This behavior illustrates the principal advantage of a meteorology-dependent upper-air closure: turbulent layers can move with the atmospheric state instead of being fixed by altitude. Because the inputs are synthetic, however, these results establish responsiveness rather than the accuracy of the predicted layer positions or amplitudes.

\begin{figure}[t]
\centering
\includegraphics[width=\columnwidth]{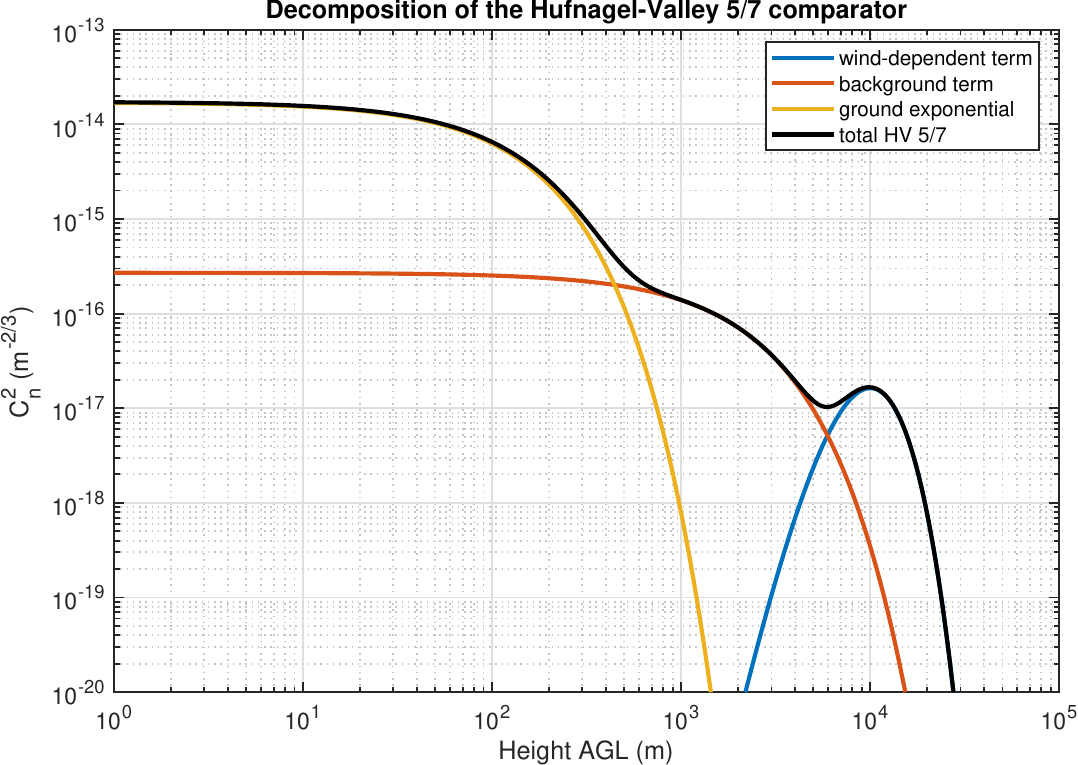}
\caption{Decomposition of the ground, intermediate, and wind-dependent terms of \ac{hv} 5/7.}
\label{fig:hvdecomposition}
\end{figure}

Figure~\ref{fig:hvdecomposition} shows that the near-ground \ac{hv} profile is controlled by the $A\exp(-z/100)$ term, while the intermediate exponential decays on a \SI{1500}{m} scale and the wind-dependent term contributes primarily in the upper atmosphere. Adjusting the \ac{hv} wind parameter therefore cannot, by itself, correct a discrepancy confined to the surface layer. This separation of roles is one reason that a meteorology-dependent replacement of the ground term must be evaluated independently from the upper-atmosphere contribution.

\begin{figure}[t]
\centering
\includegraphics[width=\columnwidth]{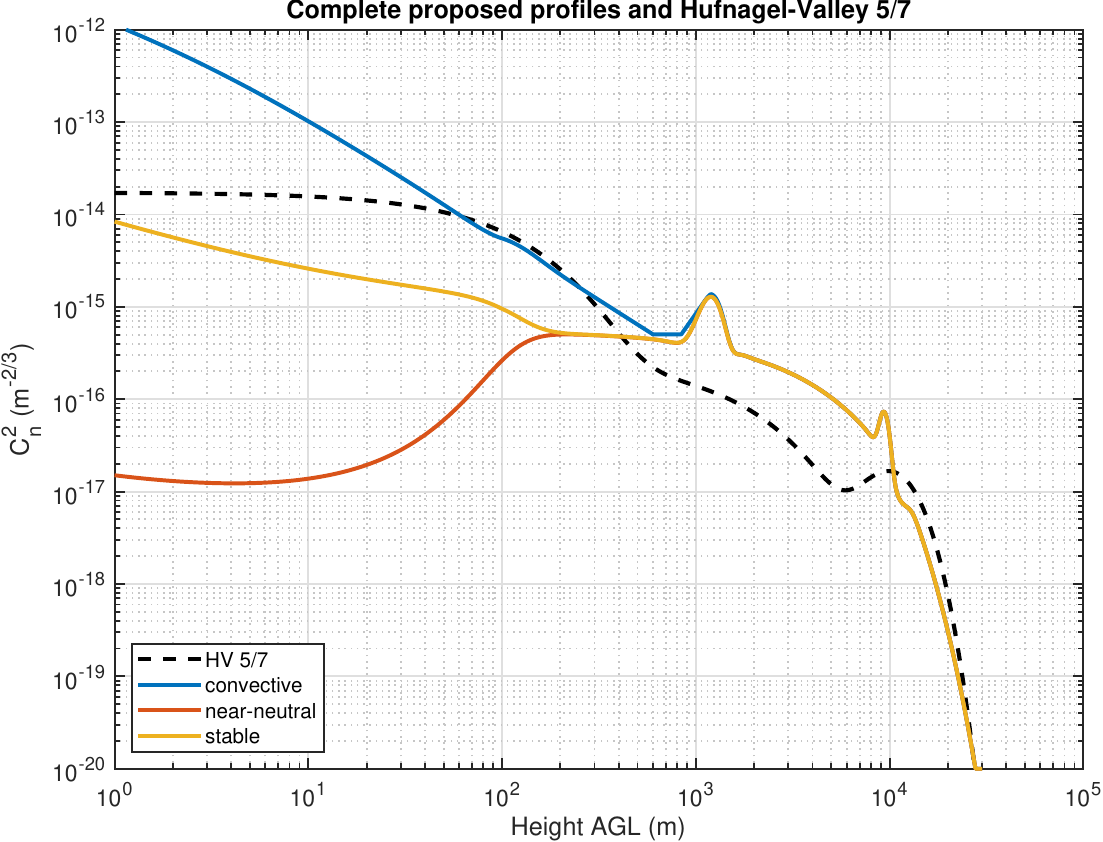}
\caption{Complete regime-dependent profiles compared with the unscaled \ac{hv} 5/7 profile.}
\label{fig:fullprofiles}
\end{figure}

\begin{figure}[t]
\centering
\includegraphics[width=\columnwidth]{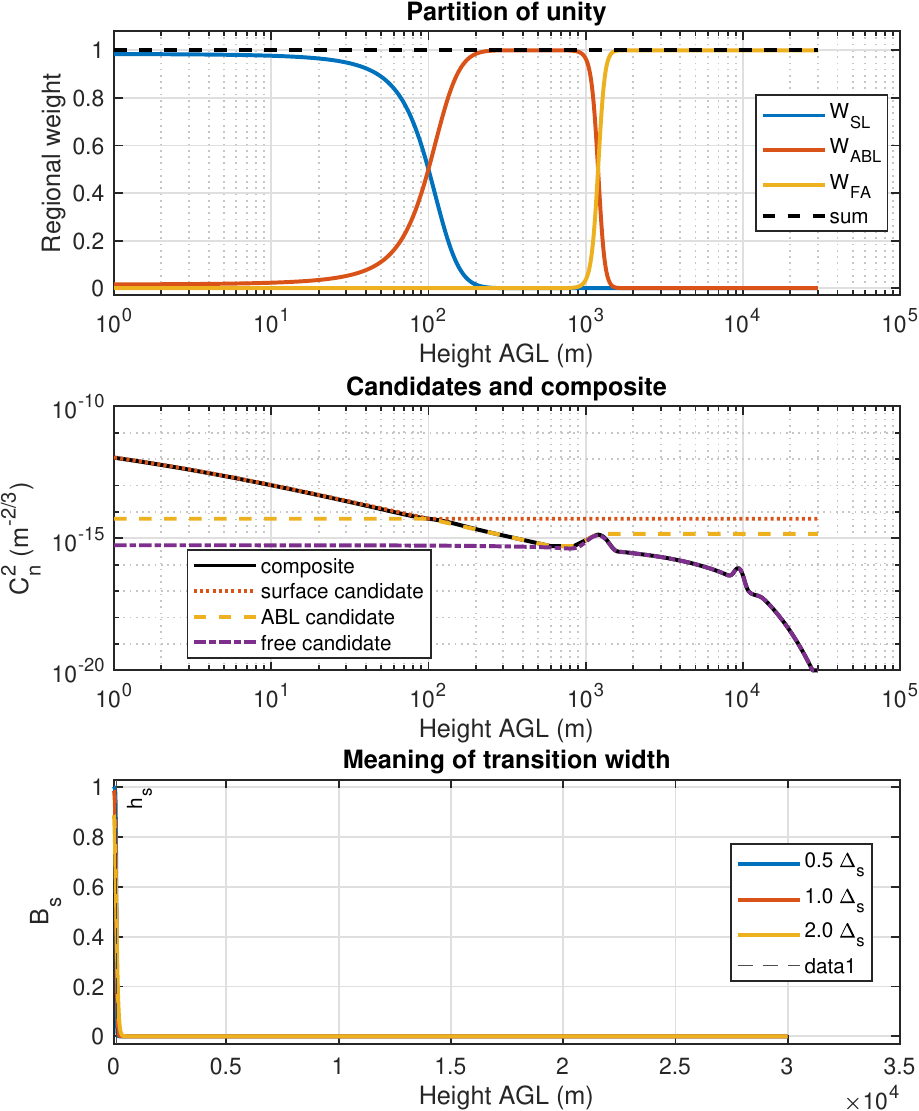}
\caption{Regional weights (left) and blended $C_n^2$ profile (right) for the convective reference case.}
\label{fig:weights}
\end{figure}

Figure~\ref{fig:fullprofiles} compares the complete regime-dependent calculations with the unscaled \ac{hv} profile. We observe the largest regime separation near the surface and within the boundary layer, where heat flux and stability differ among the prescribed cases. The profiles become more similar aloft because they share the same dissipation-based free-atmosphere input. Figure~\ref{fig:weights} verifies the numerical assembly of these components. Each candidate dominates in its intended altitude interval, the three weights sum to unity, and the logistic transitions suppress numerical jumps without adding the candidate fields as independent turbulence sources. The smoothness visible in the composite is therefore a consequence of the partition-of-unity construction rather than evidence that the physical interfaces themselves are diffuse.

For a zenith plane wave propagating downward from $H_{\rm top}$ to a receiver at ground level, the weak-fluctuation Rytov variance is \citep{andrews2005laser}
\begin{equation}
\begin{aligned}
\sigma_{R,\downarrow}^{2}
&=2.25k_0^{7/6}
\int_{0}^{H_{\rm top}} C_n^2(z)z^{5/6}\,\mathrm{d}z,
\end{aligned}
\label{eq:rytovdown}
\end{equation}
where $k_0=2\pi/\lambda$ is the optical wavenumber and $H_{\rm top}$ is the transmitter altitude above the ground receiver. In the weak-fluctuation limit, the irradiance scintillation index is approximated by
\begin{equation}
\begin{aligned}
\sigma_I^2
&=\frac{\left\langle I^2\right\rangle
-\left\langle I\right\rangle^2}
{\left\langle I\right\rangle^2}\simeq \sigma_{R,\downarrow}^{2}.
\end{aligned}
\label{eq:scintillationindex}
\end{equation}
where $I$ is received irradiance and angle brackets denote an ensemble or appropriate time average. The $z^{5/6}$ kernel in Eq.~\eqref{eq:rytovdown} is specific to the stated downlink geometry with the receiver at $z=0$; an uplink has a different path weighting.

For the direction-reversed plane-wave diagnostic, with the transmitter at ground level and the receiver at altitude $H_{\rm top}$, the corresponding uplink expression is \citep{andrews2005laser}
\begin{equation}
\begin{aligned}
\sigma_{R,\uparrow}^{2}
&=2.25k_0^{7/6}
\int_{0}^{H_{\rm top}} C_n^2(z)
\left(H_{\rm top}-z\right)^{5/6}\,\mathrm{d}z.
\end{aligned}
\label{eq:rytovup}
\end{equation}
Here $H_{\rm top}-z$ is the remaining propagation distance from a turbulent layer at height $z$ to the elevated receiver. 

\begin{figure}[t]
\centering
\includegraphics[width=\columnwidth,trim=0 0 0 18pt,clip]{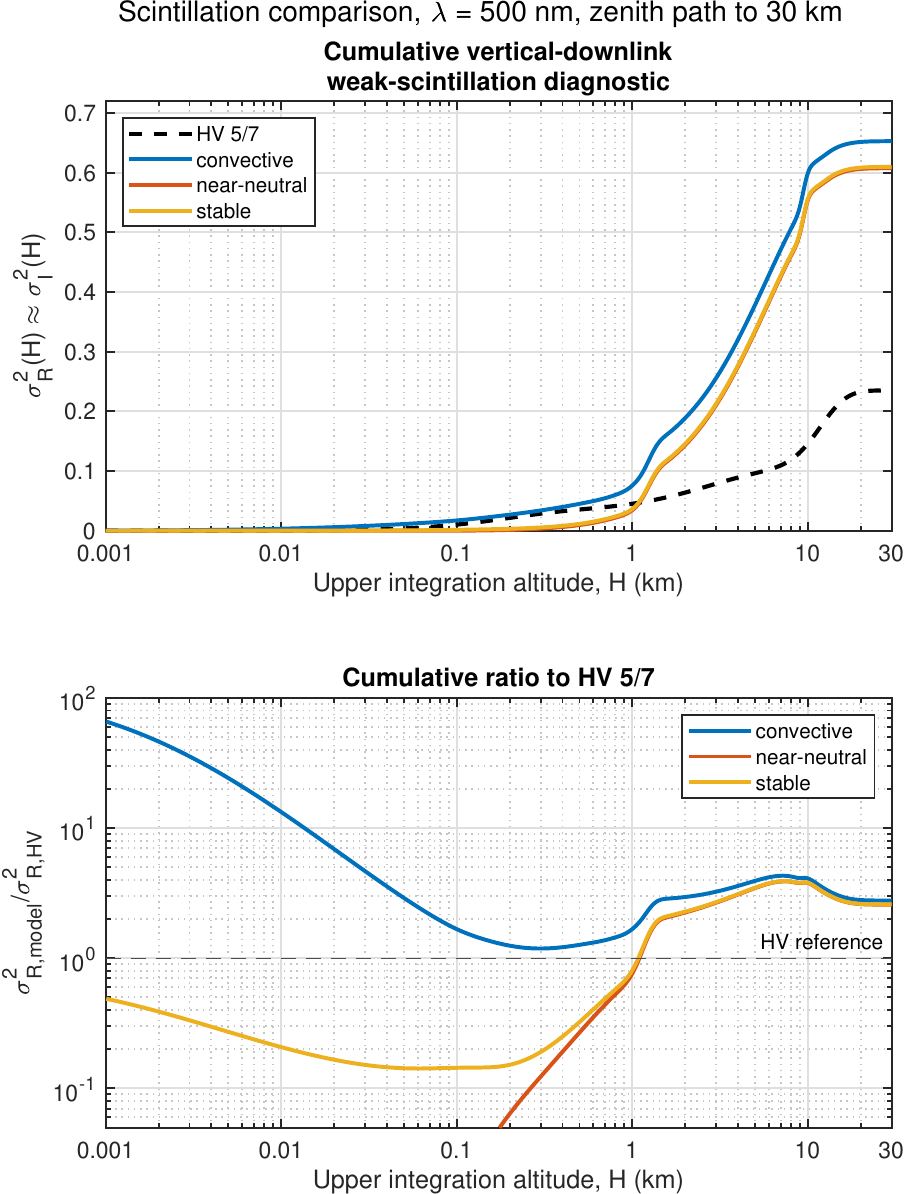}
\caption{Cumulative downlink Rytov variance (left) and ratio to \ac{hv} 5/7 (right) at \SI{500}{nm}.}
\label{fig:scintillation}
\end{figure}

Eqs.~\eqref{eq:rytovdown} and \eqref{eq:scintillationindex} are evaluated for $H_{\rm top}=\SI{30}{km}$, $\lambda=\SI{500}{nm}$, a vertical path, and a point receiver. Fig.~\ref{fig:scintillation} plots the cumulative value obtained when the upper integration altitude $H$ is progressively increased, together with the corresponding cumulative ratio to \ac{hv}. At \SI{30}{km}, \ac{hv} 5/7 gives $\sigma_R^2=0.235$, whereas the convective, near-neutral, and stable profiles give $0.653$, $0.608$, and $0.610$, respectively. Their terminal ratios to \ac{hv} are $2.78$, $2.58$, and $2.59$. Because the proposed values are not deeply within the asymptotic condition $\sigma_R^2\ll1$, they should be interpreted as first-order scintillation indicators rather than high-accuracy irradiance-variance predictions. The similar near-neutral and stable terminal values arise because their common elevated dissipation layers receive strong $z^{5/6}$ weighting, while their different near-ground values contribute comparatively little.

\begin{figure}[t]
\centering
\includegraphics[width=\columnwidth,trim=0 0 0 18pt,clip]{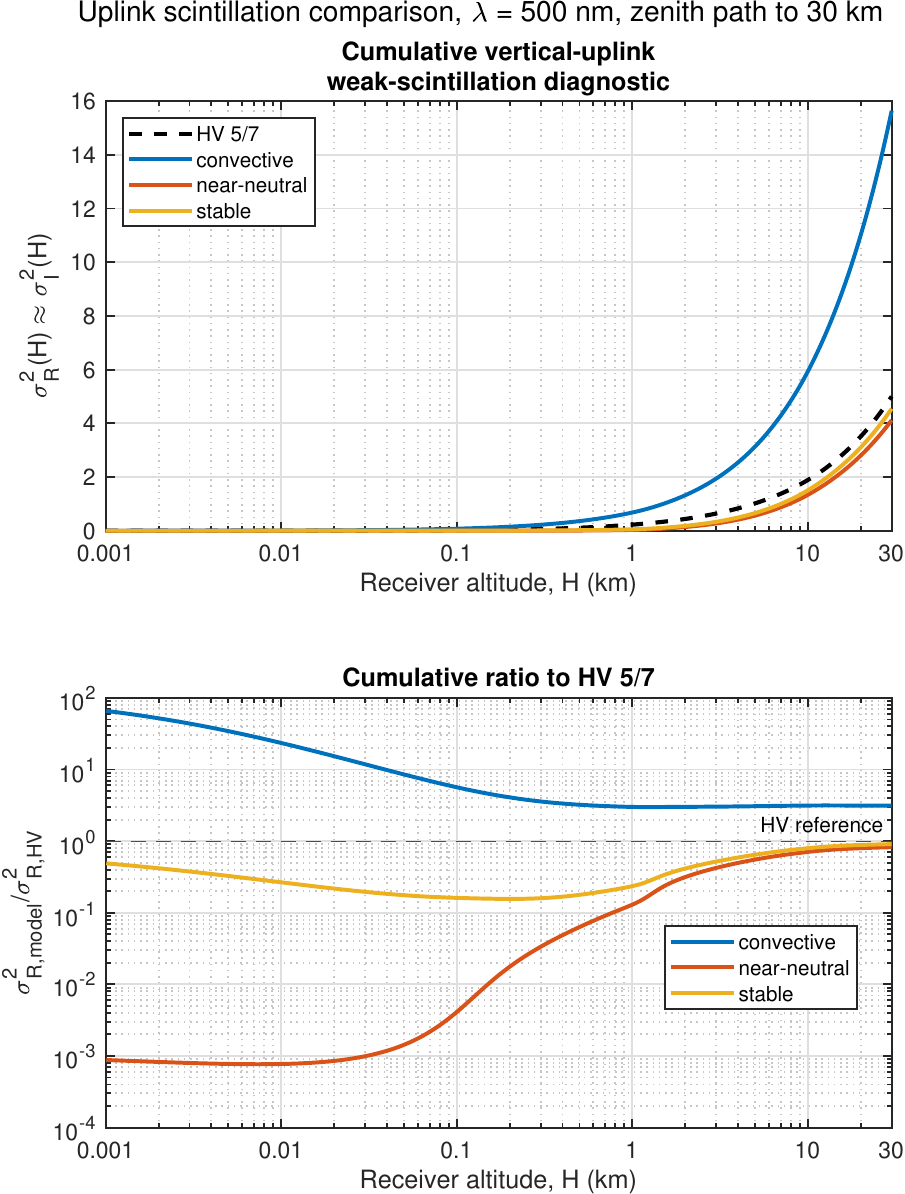}
\caption{Cumulative uplink Rytov variance (left) and ratio to \ac{hv} 5/7 (right) at \SI{500}{nm}.}
\label{fig:scintillation-uplink}
\end{figure}

Fig.~\ref{fig:scintillation-uplink} applies Eq.~\eqref{eq:rytovup} to the same prescribed profiles and wavelength. In contrast to the downlink kernel, the uplink kernel gives large leverage to turbulence near the ground because $H_{\rm top}-z$ is largest there. The comparison therefore tests the part of the proposed formulation that differs most strongly among convective, near-neutral, and stable surface states. As for the downlink calculation, the plotted values are weak-fluctuation diagnostics and not validated link-performance predictions.

The comparison with the \ac{hv} model requires a clear distinction between the intended roles of the two formulations. The \ac{hv} model provides a compact climatological representation suitable for engineering calculations, whereas the proposed model describes a turbulence profile conditioned on the prevailing meteorological state. Although this meteorological dependence provides a more detailed physical framework, greater complexity alone does not imply greater predictive accuracy. Any improvement over the \ac{hv} model must therefore be demonstrated using independent measurements collected concurrently with the required meteorological observations. Such validation should assess both the predicted \(C_n^2\) profiles and the resulting propagation quantities, including integrated turbulence strength and scintillation.

\section{Conclusion}

This study formulated a full-column, regime-dependent $\Cn$ hierarchy whose main equation is the convex composite in Eq.~\eqref{eq:composite}. Its motivation is not limited to adding meteorological variables to \ac{hv}. The broader objective is to correct altitude-dependent discrepancies that arise when a fixed climatological profile is applied to atmospheric states with different surface forcing, \ac{bl} depth, inversion structure, shear, or elevated turbulent layers. Its direct replacement for the \ac{hv} near-ground exponential is Eq.~\eqref{eq:surface}, which combines wavelength-dependent dry refractivity, a flux-derived temperature scale, aerodynamic height, and stability-dependent similarity. Strong convection is represented by a \ac{bl}-depth-dependent three-zone shape, while stable, near-neutral, and free-atmospheric turbulence use a dissipation closure when $\varepsilon$ and $\chi$ are available. The numerical results in this paper use the explicitly stated dry, visible-wavelength mode; the moist covariance equation identifies the extension required for humid applications.

The analysis establishes mathematical consistency rather than empirical accuracy. The component weights are positive and sum to unity, the limits recover the intended regional components. Deterministic prescribed-input comparisons show that meteorological conditioning can change point values by orders of magnitude, propagation-weighted integrals by factors of approximately two to three, and the stated weak-scintillation diagnostic by factors of approximately $2.6$--$2.8$ relative to \ac{hv} 5/7. 

The framework resolves several ambiguities that otherwise undermine composite models of optical turbulence. \ac{agl} and \ac{msl} coordinates are distinguished; equations taken from the literature are separated from proposed classification and blending rules; the two Tatarskii outer-scale conventions are not mixed; and numerical verification is not presented as observational validation. The next step is calibration and independent testing with coincident surface fluxes, upper-air meteorology, and vertically resolved $\Cn$ measurements. Success should mean agreement with realistic measurements across altitude and across convective, neutral, stable, transition, and free-atmospheric regimes.

\bibliographystyle{IEEEtran}

\bibliography{cas-refs}

\EOD

\end{document}